\documentclass[preprint, twocolumn,5p]{elsarticle}

\usepackage{graphicx}
\usepackage{multirow}
\usepackage{subfigure}
\usepackage{amsthm}
\usepackage{amssymb}
\usepackage{amsfonts}
\usepackage{amsmath}

\begin{document}

\title{Enhancing Information Dissemination in Dynamic Wireless Network using Stability and Beamforming}

\author[TSP]{Rachit Agarwal\corref{cor1}}
\ead{rachit.agarwal@telecom-sudparis.eu}
\author[TSP]{Vincent Gauthier}
\ead{vincent.gauthier@telecom-sudparis.eu}
\author[TSP]{Monique Becker}
\ead{monique.becker@telecom-sudparis.eu}

\cortext[cor1]{Corresponding author}
\address[TSP]{Telecom SudParis, 9 Rue Charles Fourier, Evry, 91011, France}

\begin{abstract}
Mobility causes network structures to change. In $PSN$s where underlying network structure is changing rapidly, we are interested in studying how information dissemination can be enhanced in a sparse disconnected network where nodes lack the global knowledge about the network. We use beamforming to study the enhancement in the information dissemination process. In order to identify potential beamformers and nodes to which beams should be directed we use the concept of stability. We first predict the stability of a node in the dynamic network using truncated levy walk nature of jump lengths of human mobility and then use this measure to identify beamforming nodes and the nodes to which the beams are directed. We also develop our algorithm such that it does not require any global knowledge about the network and works in a distributed manner. We also show the effect of various parameters such as number of sources, number of packets, mobility parameters, antenna parameters, type of stability used and density of the network on information dissemination in the network. We validate our findings with three validation model, no beamforming, beamforming using different stability measure and when no stability measure is associated but same number of node beamform and the selection of the beamforming nodes is random. Our simulation results show that information dissemination can be enhanced using our algorithm over other models.
\end{abstract}

\begin{keyword}
Dynamic Network, Human Mobility, Stability, Information Dissemination, Beamforming
\end{keyword}

\maketitle

\section{Introduction}\label{sec:introduction}

In dynamic communication networks such as $MANET$s, connectivity of the network has always been the point of the discussion because it affects the dissemination of the information within the network. Recent studies have shown that mobility can both speed up as well as slow down the dissemination of the information in the network \cite{Kivela2012}. The speed up of the information dissemination process is because the nodes have a higher probability of meeting other nodes in the network that either have or do not have the information. While a reduction in the rate of information dissemination process could happen as the mobility could leave the network disconnected. For example, consider a network with one large giant component, the maximum time taken to disseminate the information in the network in static case is of the order of the diameter of the network. However when the network is mobile, the giant component may break into smaller components. This disconnection would stop the transmission of the information from one component to another until the connection is made or a node with the information connects to the component that does not have the information. Other than mobility there are many other factors that also affect the information dissemination in the network. These include, bursty data \cite{Karsai2011}, strength of the tie \cite{Miritello2011}, network structure \cite{Nicosia2011}, activity pattern \cite{Vazquez2007}, content type \cite{Wang2009}, node characteristics \cite{Wang2011a}, altruism \cite{Hui2009} etc.

Contrary to $MANET$s, in a static network, \cite{Agarwal2011,Agarwal2012} showed that beamforming at the transmission end could make a disconnected wireless network connected and could help in faster information dissemination. Though the focus of their technique was to achieve small world like characteristics, their approach also used natural techniques like flocking and lateral inhibition to solve the issue of connectivity in a distributed way without the global knowledge of the network. Small world properties of a network are marked by small $A$verage $P$ath $L$ength and high $C$lustering $C$oefficient ($APL$ was defined as the mean of distance between all pairs of nodes in the network while $CC$ was defined as the fraction of number of triangles that exists to all possible triangles that could exist). Further, authors of \cite{Agarwal2011,Agarwal2012} used closeness centrality \cite{Freeman1979} to classify nodes in the network into 3 categories, 1) nodes that have low closeness centrality beamform, 2) nodes to whom beams are directed have high closeness centrality and 3) nodes that remain omnidirectional. However, in dynamic network, it is argued in \cite{Peruani2010} that beamforming can also increase the dissemination of the information. The authors of \cite{Peruani2010} showed that only small fraction of nodes are needed to beamform for enhancing the information dissemination in the mobile network. They also showed that information dissemination process depends on the width of the beam and better performance can be achieved if beams are longer and narrower if beams are created using same power as that of omnidirectional beam. They further showed that further gains are achieved if antenna rotates. However, in their algorithm they only considered single source single packet information dissemination scenario. Another work related to beamforming in $MANET$s include \cite{Li2012}. The focus of \cite{Li2012} was to study the affects of beamforming in human centric vehicular networks. The authors in \cite{Li2012} also showed that beamforming can enhance information dissemination in mobile networks with similar observations as of \cite{Peruani2010}. However the only difference between \cite{Peruani2010} and \cite{Li2012} being in the selection of nodes that beamform. Moreover, contrast to \cite{Agarwal2011,Agarwal2012}, \cite{Li2012} used directional antenna at reception.


Moreover, we think that information dissemination can can be further enhanced as compared to \cite{Peruani2010,Li2012} in terms of time in a dynamic network where connectivity is very less. The above statement forms our motivation for this paper where we discuss a mechanism to enhance information dissemination in dynamic networks. Further in this paper, we consider $P$ocket $S$witched $N$etworks, $PSN$s, a type of $MANET$, where devices are small, have limited energy and are carried by humans \cite{Hui2005a,Hui2005,Chaintreau2005}. Going back to the concept mentioned in \cite{Agarwal2011,Agarwal2012}, in dynamic networks when global data is missing, computation of the centrality measures to determine nodes that should beamform in order to reduce $APL$ in a distributed manner is highly complex and energy wasteful. Thus the algorithm in \cite{Agarwal2011,Agarwal2012} cannot be used in dynamic conditions. In \cite{Kim2012a} authors proposed a mechanism to predict centrality measures. However, their method used global data to perform prediction. Their method aggregates all the previous snapshot of the graphs to perform prediction. Many studies have argued that aggregation is not good as it results into false estimations. Towards this, we provide an efficient way to enhance information dissemination by associating a stability measure with a local centrality measure that is computed based only upon one previous snapshot and not aggregated snapshot.

A stability measure reveals how stable a system is. A system can be a network \cite{Jurman2010,Hanneke2010,Tang2010,Braha2006,Braha2009}, a link in the network \cite{Zayani2012} or a node in the network \cite{Brust2007}. Section \ref{subsec:stability} provides a brief overview of these methods. As we are interested in computing node characteristics in the dynamic network, stability of network and stability of the link interests us less. For node stability, a node is said to be stable either when it is not moving or when the node is moving but it is keeping some percentage of its interactions intact. In this paper, we would limit our focus to the second definition of node stability. Thus we use degree to compute stability of a node. Degree is a local measure and it does not require any global information. In terms of degree, if a node is $p\%$ stable it would mean that the node has maintained $p\%$ of its initial set of neighborhood. In \cite{Brust2007} authors used degree over two consecutive snapshots to compute stability of the node. In a directional and mobile scenario however, computation of the new neighborhood, then beamforming and determining again the new neighborhood would consume energy and time. Thus it would be beneficial for a node to estimate its new neighborhood and its stability beforehand. This requires a node to be able to predict its stability measure at each time instant even before knowing about the neighborhood. Towards this, the nodes calculate the probability of being connected to one of its old neighbor using the mobility characteristics of $PSN$ (jump length distribution of human mobility pattern, i.e., truncated levy walk model). Once the probabilities are calculated the node average out the probabilities to estimate its stability.

In dynamic network where information dissemination is the main issue, low stability of the node would mean that the node has met many different nodes. This would increase the probability for a node to receive the information in the network. Moreover, if that node has high degree it would imply that the node would be able to receive more information through its neighbors and in turn will be able to send out more information in the network. Further, irrespective of the degree, if a node is highly stable it would mean that the node virtually has same neighborhood and therefore has low chances of receiving different information that is flowing in the network. Coming back to our original problem of enhancing dissemination of information in the dynamic network, a low stability node becomes a prime candidate for a node that has to beamform while high stability nodes candidates for nodes toward which beams should be directed. This would allow high stability nodes to also receive information in the network. Further, all different scenarios to which nodes can be potential beamformers and recipient of the beams are mentioned in section \ref{sec:model}.

Thus, in this paper, we use above mentioned discussion to build our algorithm and focus on how information dissemination can be further enhanced in the dynamic spatial network where nodes lack global knowledge about the network. We design our algorithm such that it first addresses how nodes can predict their stability each time using previous instant information and then apply beamforming. We then use this stability measure to identify nodes that beamform and the nodes towards which the beams should be directed in order to disseminate the information in the network much faster. Towards this, in section \ref{sec:model} we provide an analytical model for predicting the stability of a node and provide a mechanism for the nodes to determine who will beamform to whom. Extensive simulations led us to formulate different scenarios where we study the impact of various parameters like different number of sources and packets, density of the network, mobility parameters, radius of communication, stability parameters and antenna parameters. We find that information dissemination greatly depends on the above mentioned parameters and have an impact in percolation of number of nodes having all the packets. This percolation shifts towards left or right depending on the parameter used.

Further, this paper is structured such that it gives a brief overview of related work and useful concepts in section \ref{sec:relatedwork}, the model outline in section \ref{sec:model} and the results obtained in section \ref{sec:simresult}. We conclude the paper by presenting conclusion of the study, section \ref{sec:conclusion}.

\begin{table}[!htb]
\centering
    \begin{tabular}{|l|l|}
        \hline
        \textbf{Notation} & \textbf{Meaning}\\
        \hline
        $\alpha$ & power law exponent constant\\
        $\beta$ & exponential cutoff constant\\
        $p()$ & probability distribution function\\
        $r_{g}$ & radius of gyration\\
        $g(\vartheta,\varphi)$ & gain\\
        $I(\vartheta,\varphi)$ & radiation intensity\\
        $r$ & transmission radius\\
        $M$ & max. antenna elements available with $v$\\
        $m$ & number of antenna elements used by $v$ to \\
            & beamform $| m\in [2,M]$\\
        $B_{b}$ & bore-sight direction\\
        $G$ & initial network of $V$ vertices and $E$ edges\\
        $V_{i}$ & a node $\in V$\\
        $t$ & time\\
        $N_{V_{i}}^{t}$ & neighborhood of $V_{i}$ at time $t$\\
        $G_{t}$ & snapshot of $G$ at time $t$\\
        $X^{t}$ & adjacency matrix of $G_{t}$\\
        $\lambda_{i}^{t}$& $i_{th}$ eigen value of $X^{t}$\\
        $\mu_{i}^{t}$& $i_{th}$ eigen value of $X^{t}$\\
        $X^{t}_{V_{i},V_{j}}$ & link status between $V_{i}$ and $V_{j}$ at $t$\\
        $S(G)$ & stability of the graph\\
        $T$ & max. time\\
        $n$ & number of words\\
        $\ell$ & max. length of word\\
        $Ent(V_{i},V_{j})$ & entropy of a link between $V_{i}$ and $V_{j}$\\
        $Ent(G_{t})$ & entropy of the Graph $G_{t}$\\
        $S(V_{i})$ & stability of the node $V_{i}$\\
        $N_{in,V_{i}}^{t}$ & in-degree neighborhood of $V_{i}$ at $t$\\
        $A$ & simulation area\\
        $S_{max}$ & max. stability threshold\\
        $S_{min}$ & min. stability threshold\\
        $a_{i}$ & jump of neighborhood node of $V_{i}$\\
        $b_{i}$ & jump of $V_{i}$\\
        $l_{i}$ & distance between initial position of\\
                & neighborhood node and the final \\
                & position of $V_{i}$\\
        $\theta_{i}$ & guarantied connectivity angle \\
        $BW_{V_{i}}$ & beam width of node $V_{i}$\\
        $BL_{V_{i}}$ & beam length of node $V_{i}$\\
        $H(x)$& set of high stability nodes in the \\
                     &direction $x$\\
        $\rho$ & density of nodes\\
        \hline
    \end{tabular}
    \caption{Notations and their meaning.}
    \label{table:N_M}
\end{table}

\section{Related Work and Useful Concepts}\label{sec:relatedwork}
    In this section, we present related work and some useful concepts. We give a brief overview of information dissemination and human mobility model and beamforming in section \ref{subsec:spreading}, \ref{subsec:mobility} and \ref{subsec:beamform} respectively. We then provide overview of stability measures in section \ref{subsec:stability}. We further formulate table \ref{table:N_M} that details all the notations used in the sections hereafter.

    \begin{figure*}
        \centering
        \mbox
        {
            \subfigure[CCDF using different $\beta$.]
            {
                \includegraphics[width = 0.5\textwidth]{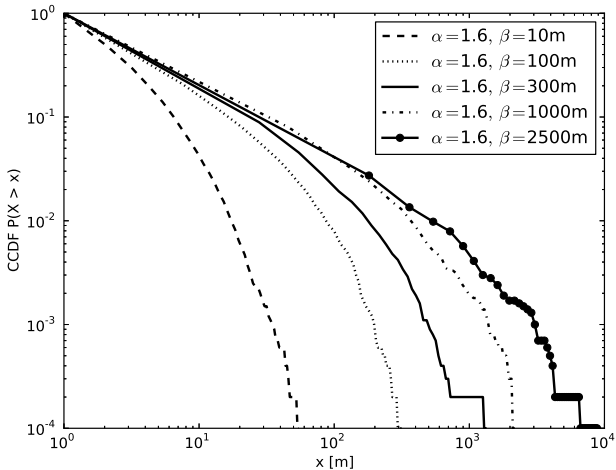}
                \label{subfig:ccdf_human_mobility_diff_beta}
            }
            \subfigure[CCDF using different $\alpha$.]
            {
                \includegraphics[width = 0.5\textwidth]{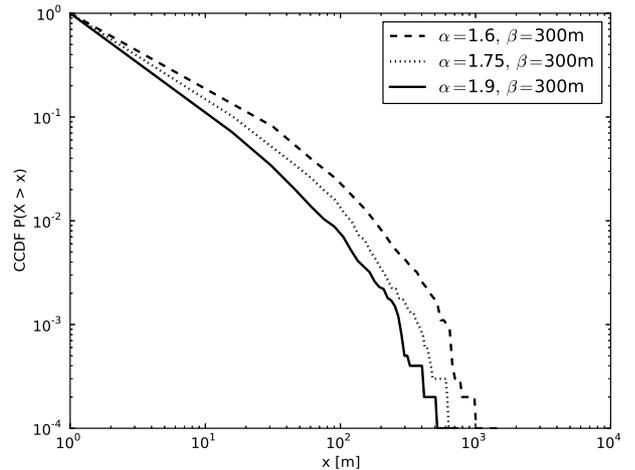}
                \label{subfig:ccdf_human_mobility_diff_alpha}
            }
        }
        \caption{Complementary Cumulative Distribution Function for human mobility using truncated power law.}
        \label{fig:pdf_human_mobility}
    \end{figure*}

    \subsection{Information Dissemination}\label{subsec:spreading}
    Models related to dissemination of information in the network have been well studied. These models are primarily inspired from disease spreading within the population. The models mainly use combinations of 4 different states, $S$usceptible, $I$nfected, $R$ecovered and $E$xposed. A node is said to be in state $S$ if it is not infected by the disease but is likely to gain the disease in future. State $I$ is when the node has the disease. State $R$ corresponds to recovered state of the node, meaning that a node was infected before but has now gained immunity while $E$ state states that a node is exposed to the disease and is highly likely to gain infection. One of the main study focusing on the information dissemination in the network was that of Watts and Strogatz's in 1998 \cite{Watts1998}. The study influenced researchers to investigate information dissemination in various different types of networks, for example, scale free networks \cite{PastorSatorras2001,Santos2005} and sparse networks \cite{Pettarin2011}. However, recently the focus has been shifted towards the study of information dissemination in dynamic networks \cite{Kivela2012,Clementi2008,Prakash2010,Schwarzkopf2010,Valler2011,Dutta2011,Clementi2011,Clementi2012}.

    In dynamic scenarios, dynamicity affects the dissemination of the information in the network \cite{Kivela2012}. It is thus critical while modeling information dissemination to know what application is targeted. For example, dissemination of a warning signal in the community should be very fast however dissemination of viruses should be very slow. Information dissemination is also affected by many other factors like, bursty data \cite{Karsai2011}, strength of the ties \cite{Miritello2011}, network structure \cite{Nicosia2011}, activity pattern \cite{Vazquez2007}, content type \cite{Wang2009}, node characteristics \cite{Wang2011a}, altruism \cite{Hui2009}, etc. In this paper, we do not concentrate on these factors and focus only on mobility. As we consider $PSN$, in the next subsection, (Cf. section \ref{subsec:mobility}), we would provide some insights to human mobility characteristics.

    \subsection{Human Mobility Model}\label{subsec:mobility}

    Many mobility models have been proposed that capture the characteristics of human mobility. A comprehensive survey of mobility properties associated with humans can be found in \cite{Karamshuk2011,Roy2011}, while a clear difference between different human mobility models can also be found in \cite{Karamshuk2011}.

    Most of the traditional models on human mobility use spatial properties of human mobility in the model. However, it was brought up that human mobility also has temporal characteristics and follows truncated power law distribution. Truncated Power Law is characterized by a probability distribution that has the properties of a power law with an exponential cutoff, eq. \ref{eq:truncatedpowerlaw} where proportionality constant is equal to $\frac{\beta^{1-\alpha}}{\Gamma(1-\alpha,\beta)}$, $\alpha$ is the power law decay exponent constant while $\beta$ is the cutoff value. Truncated power law distribution means that the distribution starts as a power law and ends as an exponential curve. The exponential cutoff is because $67\%$ of the time human movement is limited to a distance called radius of gyration, $r_{g}$ \cite{Gonzalez2008}. Other factors that influences this behavior are the boundaries assigned to the human movements and sampling of the mobility data \cite{Mossa2002}. Studies have revealed that $\alpha$ for human movement mainly lies between 1.75 $\pm$ 0.15 while $\beta$ varies \cite{Gonzalez2008}. Figure \ref{subfig:ccdf_human_mobility_diff_beta} shows the distribution of jump lengths of human movements using $\alpha=1.6$ and different $\beta$. While fig. \ref{subfig:ccdf_human_mobility_diff_alpha} shows the distribution of jump lengths of human movements using different $\alpha$ but $\beta=300.0$. The figure \ref{subfig:ccdf_human_mobility_diff_beta} show that as $\beta$ increases the jump length also increases. This is because the cutoff for large $\beta$'s is also high so the jump lengths are high. Further, figure \ref{subfig:ccdf_human_mobility_diff_alpha} show that as $\alpha$ increases jump length decreases. This is because the distribution of jump length is inversely related to $\alpha$. Further, according to this definition of truncated power law, the jump length, $x$, should be greater than or equal to $1$, ($x\geq1$). A $x<1$ would increase the probability to more than 1. Thus, for $x<1$, $p(x)$ is set to $1$.

    \begin {equation}\label{eq:truncatedpowerlaw}
        p(x)\propto x^{-\alpha}e^{\frac{-x}{\beta}}
    \end{equation}

    Other properties of human walk that obey truncated power law distribution include pause time \cite{Song2010a}, inter contact time \cite{Chaintreau2006} and $r_{g}$, \cite{Gonzalez2008}. The values of $\alpha$ and $\beta$ for pause time are 0.8 $\pm$ 0.1 and 17hr respectively. However, these values are 1.65 $\pm$ 0.15 and 350km for $r_{g}$.  We next provide a brief overview of antenna models and beamforming in section \ref{subsec:beamform}.

    \subsection{Beamforming}\label{subsec:beamform}

    When omni-directional radiations from different isotropic and equidistant antenna elements of a node interfere in a constructive and destructive manner with each other a characteristic long-range beam is achieved. The ability of a node to produce such long-range beams is called beamforming. Many different types of beamforming techniques exist in literature. Some of them are random beamforming, greedy beamforming and preferential beamforming. A brief survey of beamforming techniques used in wireless networks can be found in \cite{Agarwal2012}.

\begin{figure}[!htb]
        \centering
        \mbox
        {
            \hspace{-1.5cm}
            \subfigure[$B_{b}=0^o, 180^o$]
            {
                \includegraphics[width = 0.35\textwidth]{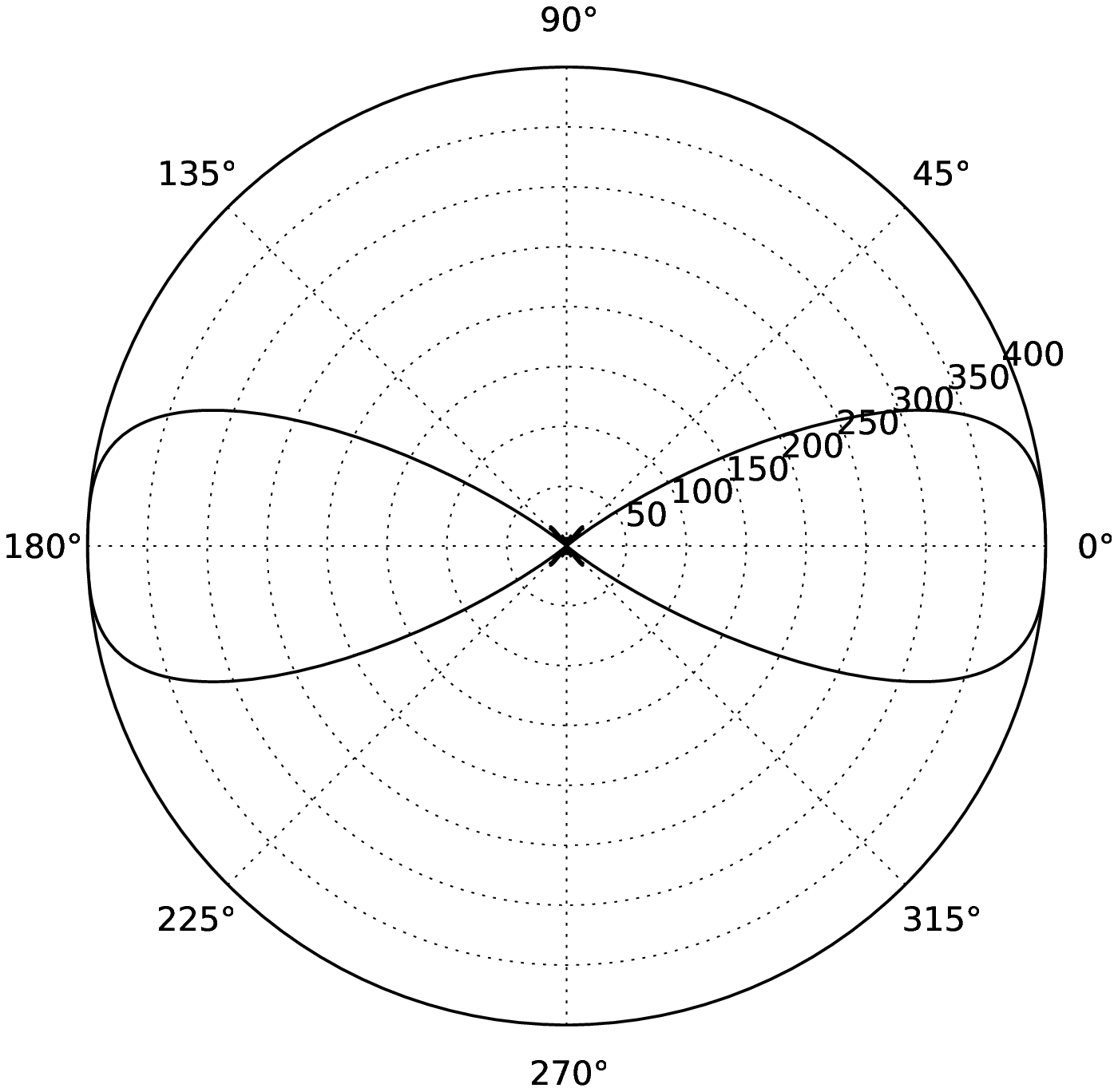}
                \label{subfig:zero}
            }
            \hspace{-2.2cm}
            \subfigure[$B_{b}=\pm 30^o$]
            {
                \includegraphics[width = 0.35\textwidth]{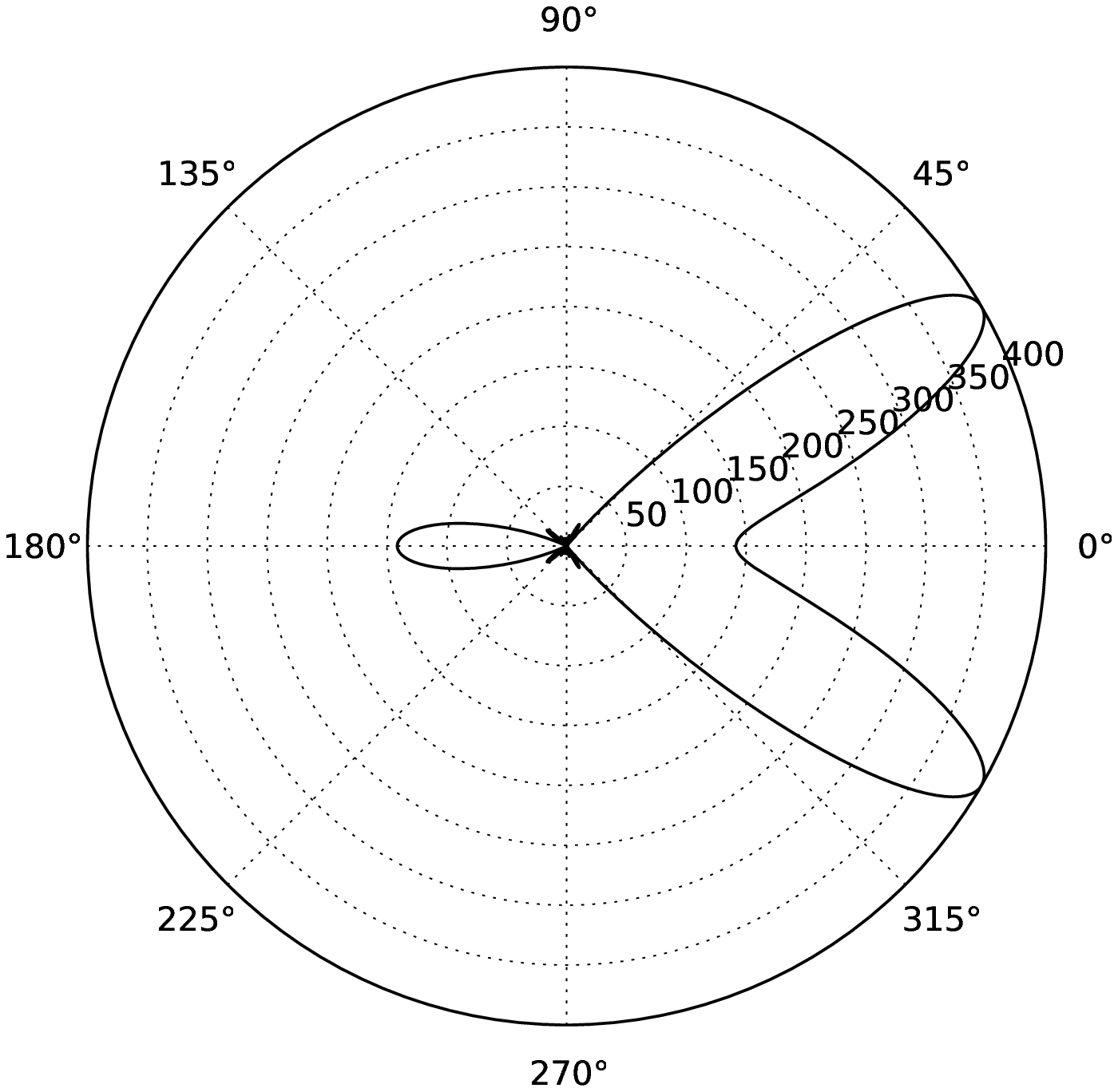}
                \label{subfig:thirty}
            }
        }
        \mbox
        {
            \hspace{-1.5cm}
            \subfigure[$B_{b}=\pm 60^o$]
            {
                \includegraphics[width = 0.35\textwidth]{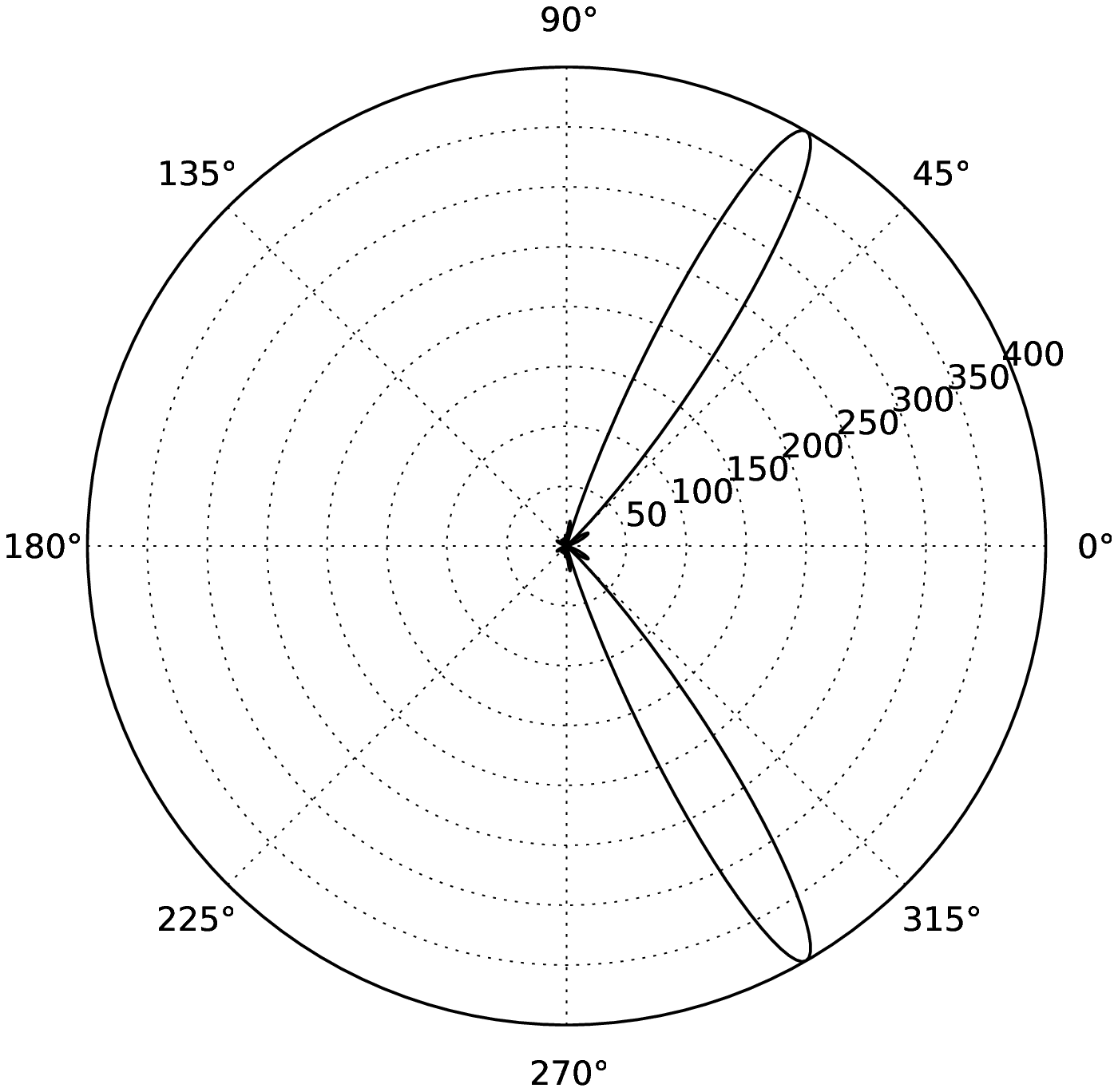}
                \label{subfig:sixty}
            }
            \hspace{-2.2cm}
            \subfigure[$B_{b}=\pm 90^o$.]
            {
                \includegraphics[width = 0.35\textwidth]{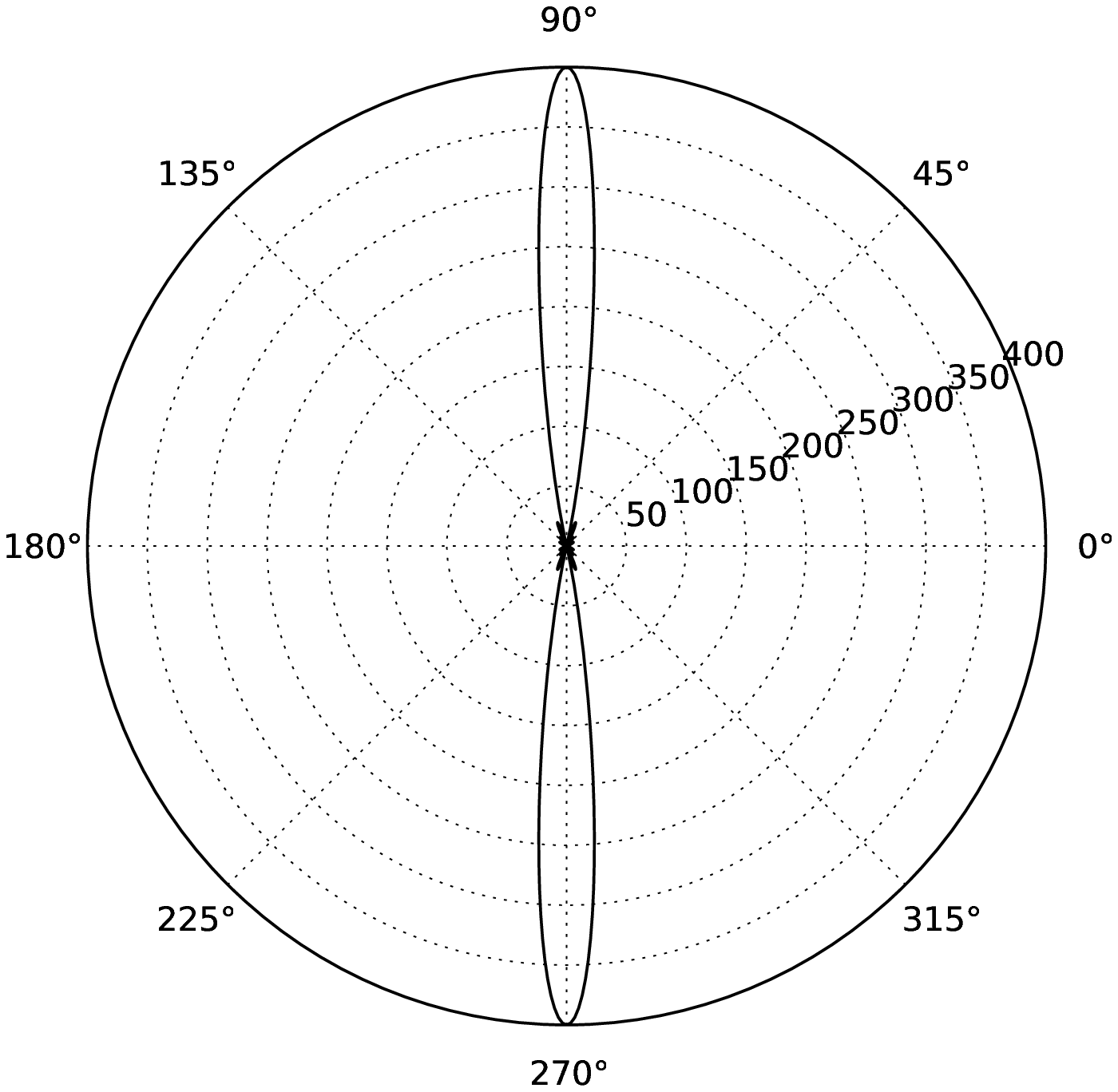}
                \label{subfig:ninty}
            }
        }
        \caption{Gain pattern obtained for different $B_{b}$, $m=8$ and $r=50$ in the $ULA$ model} \label{fig:gainpatern}
    \end{figure}

    When only single antenna element is used there is no interference and thus omni-directional radiation is maintained. However, when multiple antenna elements are used, long-range beam are obtained. The gain pattern obtained through the interference of isotropic and equidistant antenna elements is given as eq. \ref{eq:gainpattern} as defined by \cite{Balanis1997,Bettstetter2005}

    \begin{equation}\label{eq:gainpattern}
        g(\vartheta,\varphi)=\frac{I(\vartheta,\varphi)}{\frac{1}{4\pi} \int_{0}^{2\pi }\int_{0}^{\pi }u(\vartheta,\varphi) sin\vartheta d\vartheta d\varphi}
    \end{equation}
    where $\theta$ is angle with the $z$-axis, $\phi$ with the $xy$-plane and $u(\theta,\phi)$ is the radiation intensity. Research has identified many different antenna models using the above model. Among them Uniform Linear Array antenna model ($ULA$), and Uniform Circular Array antenna model ($UCA$) are most common. $ULA$ model is characterized by linear deployment of antenna elements while $UCA$ model is characterized by circular deployment of antenna elements. The gain pattern for the $ULA$ antenna model is only dependent on the number of antenna elements and has no dependency on the bore-sight direction ($B_{b}$, the direction of maximum radiation intensity). Fig. \ref{fig:gainpatern} shows the gain pattern obtained for $ULA$ model using $m=8$ and $r=50$ where $m\in M$ is the number of antenna elements used out of $M$ available antenna elements and $r$ is the transmission radius. On the other hand, for the $UCA$ antenna model, it was shown that gain pattern is dependent on both the number of antenna elements and $B_{b}$ \cite{Bettstetter2005}.

    We next provide an overview of various stability measures available in literature.

    \subsection{Stability Measure}\label{subsec:stability}

    Stability is the measure that determines how much stable a system is in a dynamic environment. Stability can be calculated by determining the percentage change in the system. Stability can be defined at different scales, for example, in dynamic graph, stability can be defined for a graph, for a node and for a link. Many different models have been proposed that can compute stability. Like, in \cite{Jurman2010} authors use spectral difference to compute the difference between two graphs. In \cite{Hanneke2010,Tang2010} authors use adjacency matrix to compute the stability of the graph. In \cite{Braha2006, Braha2009} authors rank the nodes based on degree to compute stability of the graph, in \cite{Zayani2012} authors use entropy to infer stability of the link while in \cite{Brust2007} authors define a way to compute stability of a node. Moreover, concepts from data mining can also be applied to determine the stability as suggested by Zayani et al. in \cite{Zayani2012}. Using data mining concepts, stability can be computed using similarity methods between two instances of the system into consideration. Considering a graphical system, $G(V,E)$, with $V$ nodes and $E$ edges. We explain more about different stability measures in detail in the following subsections. Table \ref{table:R_S_M} shows the differences between proposed methods.

    \begin{table}
        \centering
        \begin{tabular}{|l|l|l|}
            \hline
            \textbf{Stability} & \textbf{Method} & \textbf{Local/} \\
            \textbf{for}&&\textbf{Global}\\
            \hline
            Graph & Similarity & Global\\
            Graph & Adj. spectrum \cite{Jurman2010}& Global\\
            Graph & Adj. matrix \cite{Hanneke2010} & Global\\
            Graph & Adj. matrix \cite{Tang2010}& Global\\
            Graph & Rank Overlap \cite{Braha2006, Braha2009}& Global\\
            Link & Entropy \cite{Zayani2012}& Local\\
            Node & Neighborhood \cite{Brust2007}& Local\\
            \hline
        \end{tabular}
        \caption{Related work based on Stability measures.}
        \label{table:R_S_M}
    \end{table}

        \subsubsection{Similarity}\label{subsubsec:similarity}
        Similarity is well defined concept in data mining context. Representing a system as a vector, many similarity measures have been defined to determine the similarity between two systems. These methods include Euclidian Distance, Cosine Similarity, Jaccard Coefficient, Pearson correlation, etc \cite{LibenNowell2003,Lu2011}. A high similarity value means that two systems are almost similar and no change has occurred in the system over time. Considering a graph as a system, similarity among two instances of a graph can be computed easily. As an example, Cosine Similarity is calculated as $\frac{\sum \limits_{\forall V_{i}\in V}N_{V_{i}}^{t}\bigcap N_{V_{i}}^{t^{'}}}{|V|*|V|}$, where $N_{V_{i}}^{t*}$ is the neighborhood of $V_{i}\in V$ and $|V|$ is the number of nodes in the network at time $t*$. 


        \subsubsection{Spectral Distance}\label{subsubsec:spectraldistance}

        Change in the graphical structure can also be computed using the difference in their adjacency spectrum \cite{Jurman2010}. Adjacency spectrum is defined by the Eigen values of the adjacency matrix, $X$. In \cite{Jurman2010}, authors use distance between adjacency spectrum to compute difference in the graph structure. Further, in \cite{Jurman2010} the authors also survey different methods to compute spectral distance. We use eq. \ref{eq:SD} as defined in \cite{Pincombe2007,Jurman2010} to show difference between two graphs.

        \begin {equation}\label{eq:SD}
        SD(G_{t},G_{t'})=\begin{cases}
                        \sqrt{\frac{\sum \limits _{i=1}^{|V|}(\lambda_{i}^{t}-\mu_{i}^{t^{'}})^2}{\sum \limits _{i=1}^{|V|}(\lambda_{i}^{t})^2}} & \text{ if } \sum \limits _{i=1}^{|V|}(\lambda_{i}^{t})^2\geq \sum \limits _{i=1}^{|V|}(\mu_{i}^{t^{'}})^2 \\
                        \sqrt{\frac{\sum \limits _{i=1}^{|V|}(\lambda_{i}^{t}-\mu_{i}^{t^{'}})^2}{\sum \limits _{i=1}^{|V|}(\mu_{i}^{t^{'}})^2}} & \text{otherwise}
                        \end{cases}
        \end{equation}
        where $\lambda_{i}^{t}$ is the $i^{th}$ Eigen value of the adjacency matrix of the graph at time $t$ while $\mu_{i}^{t^{'}}$ is the $i_{th}$ Eigen value of the adjacency matrix of the graph at time $t^{'}$.

        \subsubsection{Adjacency Matrix Based Measures}\label{subsubsec:adjacencybased}

        Stability of a graph can also be measured using changes in the adjacency matrix, \cite{Hanneke2010,Tang2010}. Hanneke et al. \cite{Hanneke2010} proposed to calculate stability of the adjacency as eq. \ref{eq:SXing}. Considering two possibilities for a link existence, all possible status of a link at two given time instances can be constructed as table \ref{table:possiblinkxing}, where $X$ is the adjacency matrix and $X_{V_{i},V_{j}}^{t}$ is the status of the link between $V_{i}$ and $V_{j}$ at $t$. Hanneke et al. proposed that a link is stable if it had occurred in both the instances or did not occur while a link is not stable if it changes its state from 1 to 0 or vice versa. However, another study of Tang et al. \cite{Tang2010} said that link with status (0, 0)  in Table \ref{table:possiblinkxing} should not be considered for the stability as the link does not exist in any of the considered time instances. They proposed that stability should be computed as eq. \ref{eq:SMascolo}.

        \begin{table}
            \centering
            \begin{tabular}{|l|l|l|l|}
                \hline
                \textbf{$X_{V_{i},V_{j}}^{t}$} & \textbf{$X_{V_{i},V_{j}}^{t^{'}}$} & \textbf{Stable \cite{Hanneke2010}}& \textbf{Stable \cite{Tang2010}}\\
                \hline
                1&1&Yes&Yes\\
                1&0&No&No\\
                0&1&No&No\\
                0&0&Yes&No\\
                \hline
            \end{tabular}
            \caption{Truth Table for link between $V_{i}$,$V_{j}$.}
            \label{table:possiblinkxing}
        \end{table}

        \begin {equation}\label{eq:SXing}
        S(G)=\frac{\sum \limits _{\substack{V_{i},V_{j}\in V\\ V_{i}\neq V_{j}}} \left[X_{V_{i},V_{j}}^{t}X_{V_{i},V_{j}}^{t^{'}}+(1-X_{V_{i},V_{j}}^{t})(1-X_{V_{i},V_{j}}^{t^{'}})\right]}{|V|(|V|-1)}
        \end{equation}

        \begin {equation}\label{eq:SMascolo}
        S(G)=\frac{1}{|V|}\sum \limits _{\forall V_{i}\in V}C_{V_{i}}
        \end{equation}
        where
        \begin {equation}\label{eq:SMascolo2}
        C_{V_{i}}=\frac{1}{T-1}\sum \limits _{t}^{T-1}\frac{\sum \limits _{\forall V_{j}\in V/V_{i}}X_{V_{i},V_{j}}^{t}
        X_{V_{i},V_{j}}^{t+1}}{\sqrt{\left[\sum \limits _{\forall V_{j}\in V/V_{i}}X_{V_{i},V_{j}}^{t}\right]\left[\sum \limits _{\forall V_{j}\in V/V_{i}}X_{V_{i},V_{j}}^{t+1}\right]}}
        \end{equation}

        \subsubsection{Rank Overlap}\label{subsubsec:rankoverlap}
        Stability can also be measured by computing rank overlap as suggested in \cite{Braha2006, Braha2009}. However, done for email network, the authors suggested to compute degree of all nodes in the network at a given time instant. Once the degrees are computed, the nodes are ranked based on their degree. As the email network was large, top $\nu=1000$ nodes were selected from the ranking. This was done for different time instances and lists were made. Commonalities between two selections were then found in the form of same nodes appearing in the two lists. Mathematically overlap was model as eq. \ref{eq:rankoverlap} as shown in \cite{Hill2010}

        \begin{equation}\label{eq:rankoverlap}
        S(G) = \frac{|G_{Top}^{t}\bigcap G_{Top}^{t^{'}}|}{|V|}
        \end{equation}
        where $G_{Top}^{t}$ and $G_{Top}^{t^{'}}$ are the list of all top $\nu$ nodes at time $t$ and $t^{'}$ respectively.

\begin{figure}
            \centering
            \includegraphics[width=0.5\textwidth]{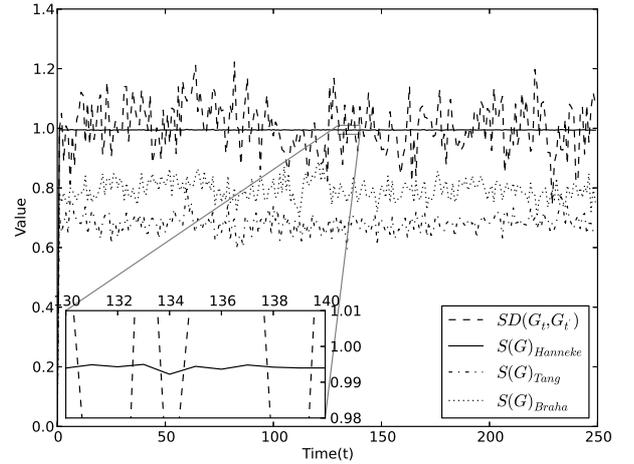}
            \caption{Relation between different stability measure values. Inset shows a zoomed-in image of stability measure computed through eq. \ref{eq:SXing}. The stability value using eq. \ref{eq:SXing} are high and $\approx0.995$ because all links that do not occur in two consecutive time instances are also considered.}
            \label{fig:allstability}
        \end{figure}

        Further, we also provide a comparison of stability measures as in section \ref{subsubsec:spectraldistance}, \ref{subsubsec:adjacencybased} and \ref{subsubsec:rankoverlap} for graphs. Figure \ref{fig:allstability}, shows differences in the stability values for a graph using different methods. The stability measures in figure \ref{fig:allstability} are calculated using two consecutive time instance graphs. Moreover, rank overlap is computed using top $50\%$ nodes. The figure shows that the stability measure differ from each other.

        \subsubsection{Entropy}\label{subsubsec:entropy}
        Entropy means the ``predictability" of the system. This means that Entropy reflects the measure of stability. If the system is lacking predictability the states of the system will not be well known and the entropy would be high. If the entropy is low it would mean that the system is predictable.

        Zayani et al. \cite{Zayani2012} used this definition to predict graphical systems. Zayani et al. used the entropy with Katz centrality to predict the occurrence of the link in the dynamic network. They argued that a link between nodes $V_{i}$ and $V_{j}$ can be represented as a series of 1's and 0's collected over time, with 1 meaning the existence of the link at a time instant and 0 meaning the absence of the link. Zayani et al's algorithm built these sequences only for those links which have occurred over time at least once. Their algorithm used learning to know about the nodes existing in the network. To compute entropy, they used an algorithm that identified maximal number of unique words, $n$, in the string of 1's and 0's. They defined entropy of the link between a pair of node, $V_{i}$ and $V_{j}$, as eq. \ref{eq:Haykel}

        \begin{figure}
            \centering
            \includegraphics[width=0.5\textwidth]{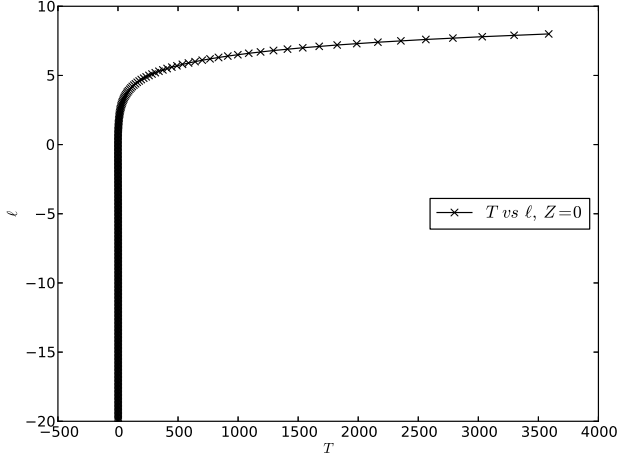}
            \caption{Variation of $\ell$ with respect to T when Z=0 for worst case sequence.}
            \label{fig:haykelequationgraph}
        \end{figure}

        \begin {equation}\label{eq:Haykel}
        Ent(V_{i},V_{j})=\frac{n*\ln(n)}{T}
        \end{equation}
        where $T$ is the time scale over which the link is observed or the maximum time. However, Zayani et al did not provide any analysis to their work. We performed a worst and best case analysis of the algorithm proposed by Zayani et al. We first identify the worst case sequence. Intuitively worst case sequence should be ``0101010101010101010101....". But, because of the periodic nature this sequence has low entropy. According to the algorithm proposed by Zayani et al, we find ``01000110110000010100111001011101110000..." sequence to be the worst case sequence. This sequence is formulated using the fact that one can use two unique words of length one (0, 1), four unique words of length two (00, 01, 10, 11), eight unique words of length three (000, 001, 010, 011, 100, 101, 110, 111) and so on. The length of this sequence can be represented using eq. \ref{eq:Haykel2}

        \begin {equation}\label{eq:Haykel2}
        T=2^{\ell+1}(\ell-1)+2+Z
        \end{equation}
        where $Z$ is the remainder of $\frac{T}{2^{\ell+1}(\ell-1)+2}$ and $\ell$ is the length of largest word. The analysis reveals that $\ell$ can be given by eq. \ref{eq:Haykel3} while $n$ in the series can be given by eq. \ref{eq:Haykel4}

        \begin {equation}\label{eq:Haykel3}
        \ell=\begin{cases}
         \frac{W(Y)+\ln2}{ln2}& \text{ if } Z=0\\
         \frac{W(Y)+\ln2}{ln2}+1& \text{otherwise}
        \end{cases}
        \end{equation}

        \begin {equation}\label{eq:Haykel4}
        n=\begin{cases}
         2^{\frac{W(Y)+2\ln2}{\ln2}}-2 & \text{if } Z=0\\
         \frac{\ln2*2^{\frac{W(Y)+3\ln2}{\ln2}}-2*(W(Y)+3\ln2)+T*\ln2}{W(Y)+2\ln2}& \text{otherwise}
        \end{cases}
        \end{equation}
        where $Y=[T/2-1]*\ln2$, $Z\in \mathbb{Z}^{*}$ and W(.) is the product log function. Fig \ref{fig:haykelequationgraph} shows the variation of $\ell$ with respect to $T$ when $Z=0$. It can be seen that the growth of $\ell$ is not so evident and it almost take $T>3586$ to have $\ell\geq8$. For $T<3$, $\ell<0$, this is contradictory to the definition of $\ell$ as $\ell$ is the word length which cannot be less than zero. Thus to compute entropy $T$ should be greater than or equal to $3$ ($T\geq3$). The best case sequence on the other hand is a sequence of all 1's or all 0's. This will lead to eq. \ref{eq:Haykel5}

        \begin {equation}\label{eq:Haykel5}
        T=\frac{\ell(\ell+1)}{2}+Z
        \end{equation}
        where $Z\in[0,\ell]$. Here $n=\ell$ when $Z=0$ and $n=\ell +1$ when $Z\neq0$. $\ell$ is further defined as eq. \ref{eq:Haykel6}

        \begin {equation}\label{eq:Haykel6}
        \ell=\frac{1}{2}*(\pm\sqrt{8T}-1)
        \end{equation}

        \begin{figure}
            \centering
            \includegraphics[width=0.5\textwidth]{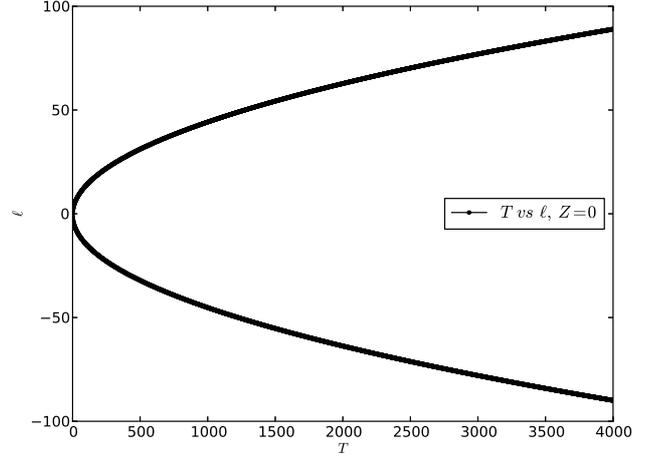}
            \caption{Variation of $\ell$ with respect to T when Z=0 for best case sequence.}
            \label{fig:haykelequationgraph2}
        \end{figure}

        Variation of $\ell$ with respect to $T$ for eq. \ref{eq:Haykel5} is shown in fig \ref{fig:haykelequationgraph2}. As $\ell$ cannot be negative we just consider $\ell\geq 0$. If above analysis is used, it will reduce the computation of words every time as done in the Zayani et al. algorithm and can save lot of node's energy. A similar method to that of Zayani et al was proposed in \cite{Song2010} where authors used location preferences instead of links.

        Studies have also revealed that entropy of a graph can also be defined. Various models, \cite{Shetty2005,Lu2008,Volchenkov2007, Riis2007,LinHan2011,Zhao2011}, have been developed to compute entropy of the graph. A brief survey of entropy measures could be found in \cite{Zayani2011, Dehmer2011}. Some of above mentioned models, \cite{LinHan2011}, use Laplacian spectrum of the graph to compute the entropy of the graph as eq. \ref{eq:entropyhancock}.

        \begin{align}\label{eq:entropyhancock}
         Ent(G_{t}) &= \sum_{i}^{|V|}(\lambda_{i}^{t}/2)\log(\lambda_{i}^{t}/2)\\ \nonumber
         &\approx \sum_{i}^{|V|}(\lambda_{i}^{t}/2)*(1-\lambda_{i}^{t}/2)
        \end{align}
        where $\lambda_{i}^{t}$ is the $i^{th}$ Laplacian Eigen value.

        \subsubsection{Neighborhood Based}\label{subsubsec:neighborhood based}
        All the above mentioned measures, except link entropy, for computing stability require global knowledge. A simple measure for computing node stability in terms of neighborhood is defined as eq. \ref{eq:nodestabilityone}

        \begin{equation}\label{eq:nodestabilityone}
        S(V_{i}) = \frac{|N_{in,V_{i}}^{t}\bigcap N_{in,V_{i}}^{t^{'}}|}{|N_{in,V_{i}}^{t}|}
        \end{equation}
        where $N_{in,V_{i}}^{t}$ is the in-degree neighborhood of $V_{i}\in V$ at time $t$. The eq. \ref{eq:nodestabilityone} identifies all those nodes that were neighbors of the node at time $t$ and $t^{'}$ and computes the ratio with the actual number of neighbor nodes at $t$. A similar method was proposed in \cite{Brust2007}. The authors of \cite{Brust2007} claim to propose a method for the computation of stability of the node but the model lacked clarity in the formulation.

        The above model captures the local computation of stability measure for the node but requires two snapshots of graphs in time, $t$ and $t^{'}$. This would mean that the node would first require the knowledge about its new neighborhood at time $t^{'}$. It would then compute stability, then beamform and then recompute the new neighborhood. The node would thus require time in performing above four steps. It is highly possible that by that time the node changes its position. To reduce computation time it would be better if the nodes are able to predict stability before. In the next section (Cf. section \ref{sec:model}) we propose a stability measure that can be computed using local neighborhood and does not require new neighborhood for the calculation of stability.

\section{Model}\label{sec:model}
In this section we provide a method to estimate the stability of the node in the dynamic environment and how this stability could be used toward enhancing the information dissemination in a $PSN$. We use truncated levy walk characteristics of human mobility to build our model. Now, let us first consider a given spatial distribution of wireless nodes in area $A$ with transmission radius $r$. This will constitute a spatial network, $G(V,E)$, with $V$ nodes and $E$ edges. Let us also assume that nodes are moving and follow truncated power law distribution of jump lengths. When nodes are moving, the network structure is constantly changing and affects the connection setup of the nodes with the neighbors. Assuming nodes to have complete information about the network is not good. Thus, let us also assume that nodes lack global information and only have local information. Further, let us also assume that mobility speed is very much less than that of transmission speed so that when data is being transmitted the node do not change their position. We provide nodes in the network with two measures, maximum stability threshold, $S_{max}$, and minimum stability threshold, $S_{min}$ and equip each node with $M$ antenna elements. The nodes use single antenna element for the omnidirectional beam, however, uses $m\in[2,M]$ antenna elements when beamforming. We consider only directional transmission for now and do not consider directional reception. Further, we use similar antenna model to perform beamforming as used in \cite{Agarwal2012}.

Absence of global information will leave nodes to have only certain information about the other nodes. Due to mobility, we limit this information to node's initial location, node's final position after the jump and the initial position of the initial neighborhood nodes. After nodes have moved, a node will have new set of neighbors that now will act as the initial neighborhood for predicting the stability of the node for the next time instant.

Following the assumptions, we build an analytical model for computation of node stability. Let the distance moved by the node $V_{i}\in V$ in $T=t^{'}-t$ be equal to $b_{i}$ and the distance between a neighbor, $V_{j}\in V$ at $T=t$, ($V_{j}^{t}$), and $V_{i}$ at $T=t^{'}$, ($V_{i}^{t^{'}}$) be $l_{i}$ where $t<t^{'}$.

\begin{figure}
    \centering
    \includegraphics[width=0.5\textwidth]{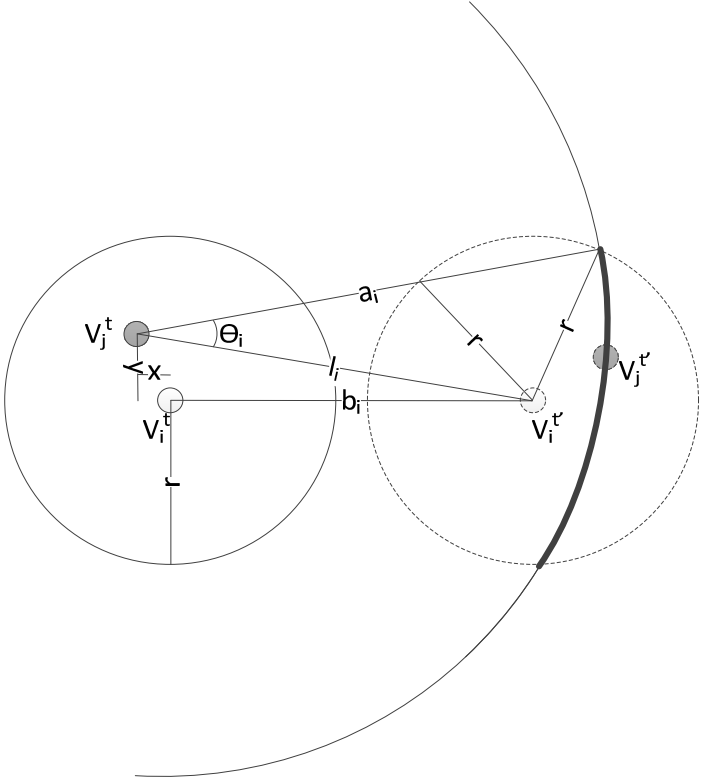}
    \caption{The probability of having a neighborhood node connected in the next time instance is given by the relative size of the dark bold line.}
    \label{fig:probabilityofconnection}
\end{figure}

In order to build the model we first assume that the node $V_{i}$ knows how much distance, $a_{i}$, its neighbor $V_{j}$ has moved in $T=t^{'}-t$. We will then relax this condition. The two nodes, $V_{i}$ and $V_{j}$, will be connected only if the node $V_{j}$ moves such that its movement is within $2\theta_{i}$ and in the direction of movement of $V_{i}$ as shown in Fig. \ref{fig:probabilityofconnection}. Using law of cosine, $\theta_{i}$ can be computed as eq. \ref{eq:modelequation1}

\begin {equation}\label{eq:modelequation1}
\theta_{i}=cos^{-1}\left(\frac{a_{i}^{2}+l_{i}^{2}-r^{2}}{2a_{i}l_{i}} \right)
\end{equation}
The probability of connection between $V_{i}$ and $V_{j}$ therefore can be given by, eq.\ref{eq:modelequation2}

\begin {equation}\label{eq:modelequation2}
P_{V_{i},V_{j}}(C|a_{i},l_{i},b_{i})=\frac{\theta_{i}}{\pi}
\end{equation}
where C stands for connection. We now remove the conditional of knowing $a_{i}$. In order to remove the conditional of knowing $a_{i}$, we need the distribution of $a_{i}$. Due to the fact that jump lengths are distributed according to truncated power law, this fact can be used to remove the conditional of knowing $a_{i}$. This results into modification of eq. \ref{eq:modelequation2} to eq. \ref{eq:modelequation3}

\begin {equation}\label{eq:modelequation3}
P_{V_{i},V_{j}}(C|l_{i},b_{i})=\int \limits _{|l_{i}-r|}^{l_{i}+r} f_{A}(a_{i})*P_{V_{i},V_{j}}(C|a_{i},l_{i},b_{i})
\end{equation}
where $f_{A}(a_{i})$ is the probability distribution function of $a_{i}$ given by eq. \ref{eq:truncatedpowerlaw}. The lower bound of the integral limit, $|l_{i}-r|$, is absolute value because if the jump of node $V_{i}$ is less than $r$, $b_{i}<r$, then the lower limit would be $r-l_{i}$, however, if jump length is larger than $r$, $b_{i}>r$, the limit would be $l_{i}-r$. This fact is captured well by the absolute value. Moreover, because the jumps cannot be negative, i.e. $b_{i}<0$, absolute value helps in addressing this fact. As discussed in section \ref{subsec:mobility}, jump lengths should be grater larger than $1$ unit. As the units are not defined, for a jump length less than $1$, the scale should be changed accordingly such that jump length is larger than $1$. Also, here $l_{i}$ should be larger than $0$. If $l_{i}=0$ then cosine term will not be defined and the probability will become 0. This fact is rather obvious as $V_{i}$ can not move to the location of the $V_{j}$ as it thinks that $V_{j}$ is already occupying the same location.

\begin{figure}
    \centering
    \mbox
    {
        \subfigure[Variation of $g(a)$ with respect to $l$ and $r$]
        {
            \includegraphics[width = 0.5\textwidth]{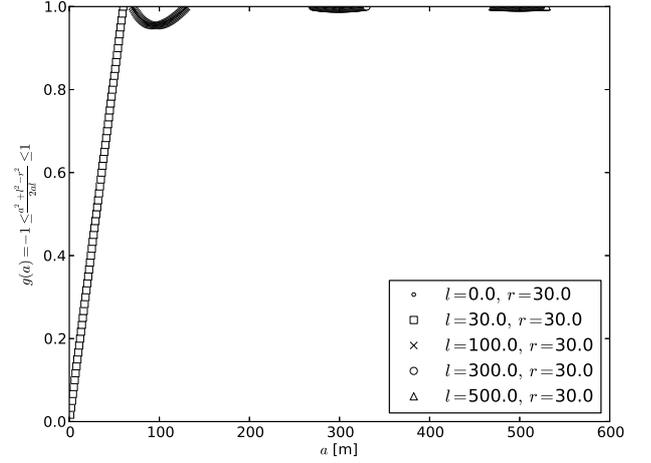}
            \label{subfig:inside_arccos}
        }
    }
    \mbox
    {
        \subfigure[Variation of $\theta$ with respect to $l$ and $r$]
        {
            \includegraphics[width = 0.5\textwidth]{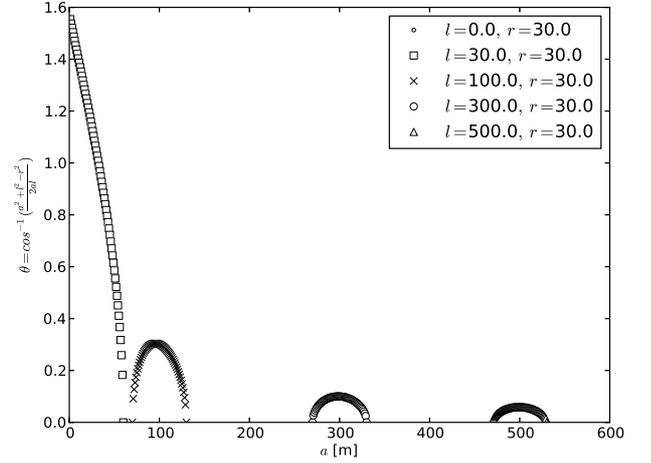}
            \label{subfig:arccos}
        }
    }
    \mbox
    {
        \subfigure[Variation of $f(a)$ with respect to $l$ and $r$]
        {
            \includegraphics[width = 0.5\textwidth]{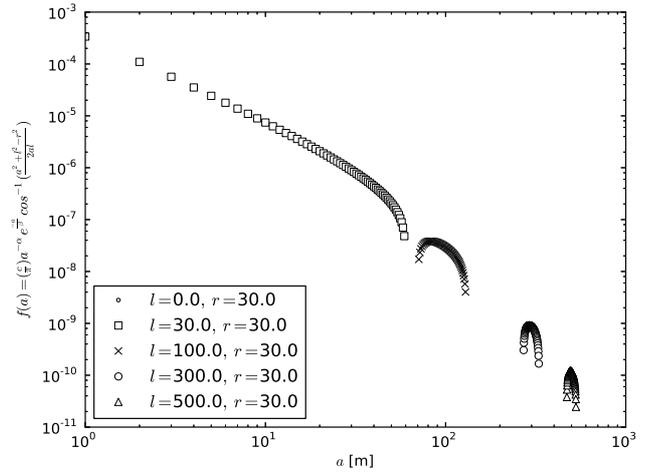}
            \label{subfig:fx}
        }
    }
    \caption{Functional analysis with varying $l$ with $\beta=300$ and $\alpha=1.6$. All distances are in meters. For $l=0$, no graph is displayed because $cos^{-1}$ is not defined at $l=0$.}
    \label{fig:analysisprob}
\end{figure}

Considering $g(a)=\frac{a_{i}^{2}+l_{i}^{2}-r^{2}}{2a_{i}l_{i}}$, $\theta=cos^{-1}(g(a))$ and $f(a)=f_{A}(a_{i})*P_{V_{i},V_{j}}(C|a_{i},l_{i},b_{i})$, figure \ref{subfig:inside_arccos} shows the variation of possible values of $g(a)$. Fig. \ref{subfig:arccos} shows the variation of $\theta$ for different values of $l_{i}$ while fig. \ref{subfig:fx} shows the variation of $f(a)$ on loglog scale for different values of $l_{i}$ and $r$ with $\beta=300m$ and $\alpha=1.6$.

As eq. \ref{eq:modelequation3} refers to the probability of keeping the connection in the next time instant, this probability directly relates to the probability of link occurrence in the next time instant. The same probability also relates to how much stable a connection is. As there can be $N_{in,V_{i}}^{t}$ in-degree neighbors of $V_{i}$ at time $t$, the total stability of the node should thus be given by eq. \ref{eq:modelequation4}

\begin {equation}\label{eq:modelequation4}
S(V_{i})=\frac{\sum \limits _{V_{j}\in N_{in,V_{i}}^{t}}P_{V_{i},V_{j}}(C|l_{i},b_{i})}{|N_{in,V_{i}}^{t}|}
\end{equation}

The probability in the eq. \ref{eq:modelequation3} is calculated for a link. This probability can be very high meaning the link is highly likely to exist, ie., highly stable or can be very low meaning the link is unstable. Summing of the probabilities provides the expected value of the number of nodes which will still be connected to the node at the next step. Here, a large number of highly stable links make the overall node stability high. For example, consider a node $V_{i}$ with in-degree neighborhood at time $t$ as $N_{in,V_{i}}^{t}=4$. Further, consider 3 link probabilities among 4 calculated as very high and one as very very low (tending towards 0). As the 3 links are highly likely the overall stability of the node $V_{i}$ at $t^{'}$ should be high. This fact is captured well through eq. \ref{eq:modelequation4}.

We use this measure of stability to identify nodes that are potential beamformers. If the value from eq. \ref{eq:modelequation4} is less than $S_{min}$, the node switches its beam from omnidirectional to directional beam. As discussed in the introduction section, the node with low stability is assumed to have met more nodes and is likely to have more information. When such a node beamforms to more stable node, it is likely to enhance the information dissemination mechanism in the network. However, when a node, $V_{i}$, at $t$ has $|N_{in,V_{i}}^{t}|=0$ but has $|N_{in,V_{i}}^{t^{'}}|>0$ at $t^{'}$, the $S(V_{i})=0$. This would mean that it is highly likely that the node does not have any information to transmit. It is thus also beneficial for another node with low stability to beamform to such nodes. Further, when $V_{i}$ at $t$ has $|N_{in,V_{i}}^{t}|=0$ and also has $N_{in,V_{i}}^{t^{'}}|=0$ at $t^{'}$, the $S(V_{i})=0$. This would also mean that the node is isolated and might not have any information to transmit. It would be beneficial to beamform to this node so that this node is not left without the information. As this node can also be the source of information, to transmit its information this node must beamform to any node in the network. To distinguish between above cases, it is necessary to associate the stability measure with the degree of the node. We thus, formulate table \ref{table:beamformer} that lists all possible combinations of stability and degree measures for the nodes that would be potential beamformers and the nodes to which these beams are to be directed.

\begin{table}
    \centering
    \begin{tabular}{|l|l|l|l|}
        \hline
        \multicolumn{2}{|c|}{\textbf{Beamform From}} & \multicolumn{2}{|c|}{\textbf{Beamform To}}\\
        \hline
        \textbf{Stability} & \textbf{Degree} & \textbf{Stability} & \textbf{Degree}\\
        \hline
        Low & Low & High & Low\\
        Low & Low & High & High\\
        Low & Low & Zero & Low\\
        Low & Low & Zero & High\\
        Low & Low & Zero & Zero\\

        Low & High & High & Low\\
        Low & High & High & High\\
        Low & High & Zero & Low\\
        Low & High & Zero & High\\
        Low & High & Zero & Zero\\

        Zero & Zero & Any & Any\\
        \hline
    \end{tabular}
    \caption{Beamforming strategies.}
    \label{table:beamformer}
\end{table}
We next describe how nodes that are beamforming can determine the angle, width and the length of the beam. We use the same adaptation of antenna configuration in our model as used in \cite{Agarwal2011,Agarwal2012} and use

\begin{equation}\label{eq:blbw}
    BW_{V_{i}}=\frac{2\pi r^2}{BL_{V_{i}}^{2}}
\end{equation}
where $BL_{V_{i}}=m*r$ as described for theoretical antenna model (Sector model) in \cite{Agarwal2012} and $BW_{V_{i}}$ is the beam width. The equation eq. \ref{eq:blbw} can provide the width of the beam but not the direction of the beam and how to determine the best $m$. As in \cite{Agarwal2011,Agarwal2012} we also use same approach in determining the length and the direction of the beam. When a node determines that it is a potential beamformer, it randomly chooses $m\in[2,M]$. As omnidirectional beam is converted to directional beam, the directional beam has to sweep $\frac{2\pi}{BW_{V_{i}}}$ sectors to cover all the directions. The node similar to \cite{Agarwal2011,Agarwal2012} sweeps all these sectors to determine the best sector. The best sector is then determined by 1) number of high stability nodes and 2) their stability value. Eq. \ref{eq:bd} models the same.

\begin{equation}\label{eq:bd}
    BD_{V_{i}}=\max_{}\left(\sum_{k\in H(BD_{V_{i}}^{j})}S(k)\right)
\end{equation}
where $H(BD_{V_{i}}^{j})$ is the set of all high stability nodes in the direction $BD_{V_{i}}^{j}$ and $j$ is a sector $\in [1,\frac{2\pi}{BW_{V_{i}}}]$. Once the direction is determined we use Sector model to beamform. The high stability nodes that were one hop away when the beamforming node was not beamforming will affect our algorithm. Directing beams to these nodes will not reduce the path length but might reduce the neighborhood. Thus, we also condition that the high stability node must be at least two hops away when the node is beamforming. If the node is not able to find any highly stable two hops away nodes within its beam length then it reverts back to omnidirectional case.

Further, beamforming makes the network asymmetric. A node will be able to transmit the data but will not be able to receive the $ACK$ for the transmission made and will not be able to know if there is a high stability node in the region of directional beam. As we are just concerned about the dissemination of the information in a broadcast medium, the problem of asymmetry is of less importance. For the second problem, again same as \cite{Agarwal2011,Agarwal2012}, we assume that a highly stable node, just to let the beamforming node know about the connection setup, creates the beam in the direction of the beamforming node and then reverts back to omnidirectional scenario.

\section{Simulation and Results}\label{sec:simresult}

We use a simulation area of $A=500m$x$500m$ to simulate our algorithm. The range of average density, $\rho$, of nodes per unit area is set to $1$x$10^{-3}$. Initially, each node operates in omnidirectional mode using $m=1$ antenna element with the omnidirectional radius as $r=30m$. We set the maximum number of antenna elements that the nodes are equipped with to $M=6$. The separation between two antenna elements computed using $WiFi$ frequency, $f=2.4Ghz$. We set $\alpha=1.6$, $\beta=300m$, $S_{min}=1$x$10^{-6}$, $S_{max}=0.7$. As $ULA$ model is more realistic we use $ULA$ antenna model to simulate our model. Through our simulations, we explore the effect time taken to disseminate the information in the network. We use Python to simulate our algorithm and use a confidence interval of $95\%$. We average all the results over $50$ initial topologies for a $T=100$. We validate our results with the scenario that assumes global knowledge, random beamforming and no beamforming. We simulate different information dissemination scenarios using simple $S$usceptible-$I$nfected model and its variations. Moreover, we show the difference in the time taken to disseminate the information in the network for multiple source multiple packet for $ULA$ antenna model and the sector model and also show the affect of various other parameters like, mobility parameters, antenna parameters and stability parameters on the information dissemination in the network.

    \subsection{Results}\label{subsec:result}
    We simulate six different scenarios for information dissemination, single source having single packet to transmit, single source having multiple packets to transmit, single source sending update information, multiple sources with each having single packet to transmit, multiple sources with each having multiple packets to transmit and multiple sources with multiple packets where sources can join over time. As a validation scheme, we compare our model results with three cases, namely, when there is no beamforming, when stability measure as proposed in section \ref{subsubsec:neighborhood based} is used and when same number of nodes as found by our model beamform but the selection of the nodes is random.

\begin{figure*}[!htb]
        \centering
        \mbox
        {
            \hspace{-0.2cm}
            \subfigure[\% nodes having packets when only one node acts as a source and it has only single packet to transmit.]
            {
                \includegraphics[width = 0.45\textwidth]{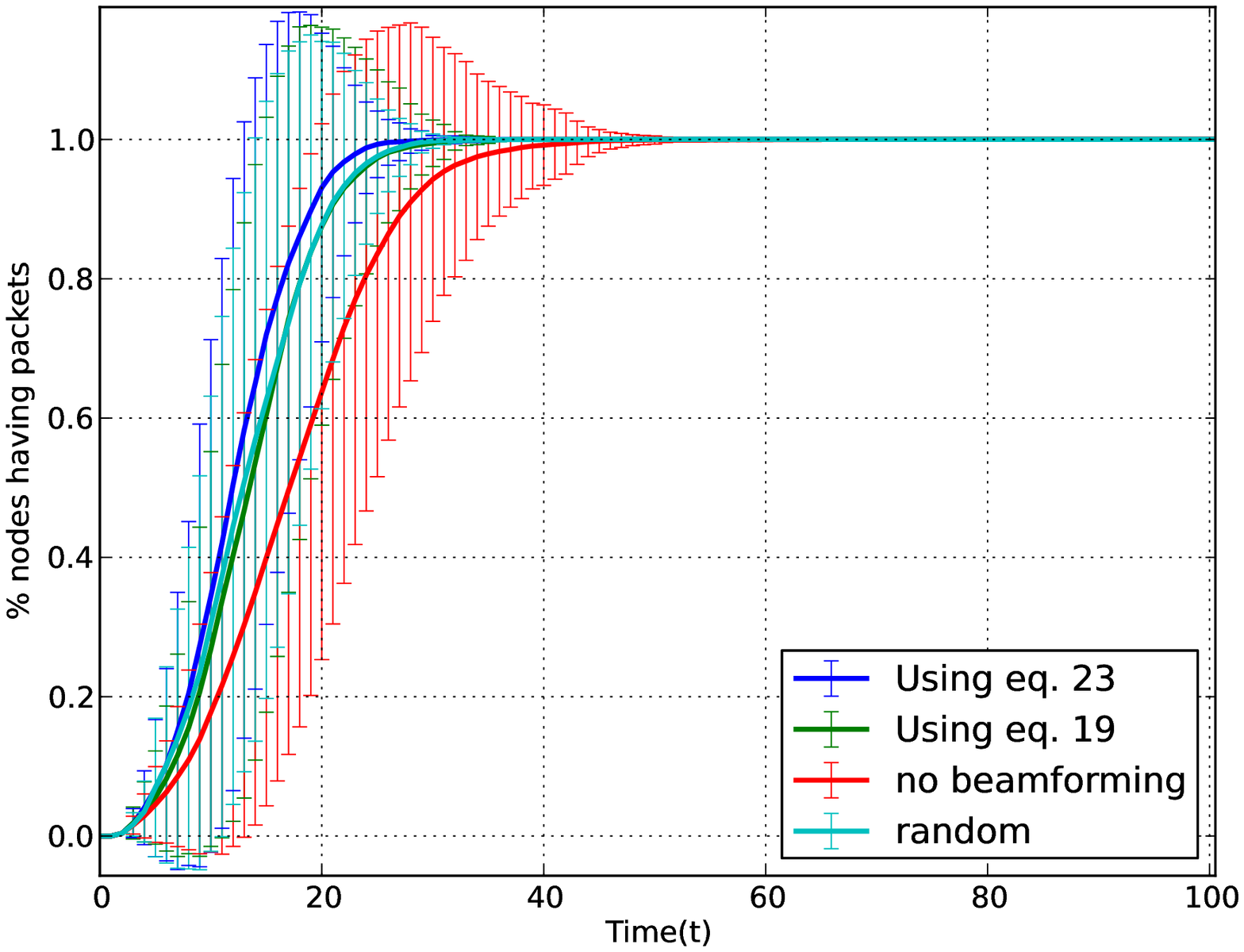}
                \label{subfig:numpacket1}
            }
            \hspace{1cm}
            \subfigure[\% nodes having packets when only one node acts as a source and it has updated packet to transmit.]
            {
                \includegraphics[width = 0.45\textwidth]{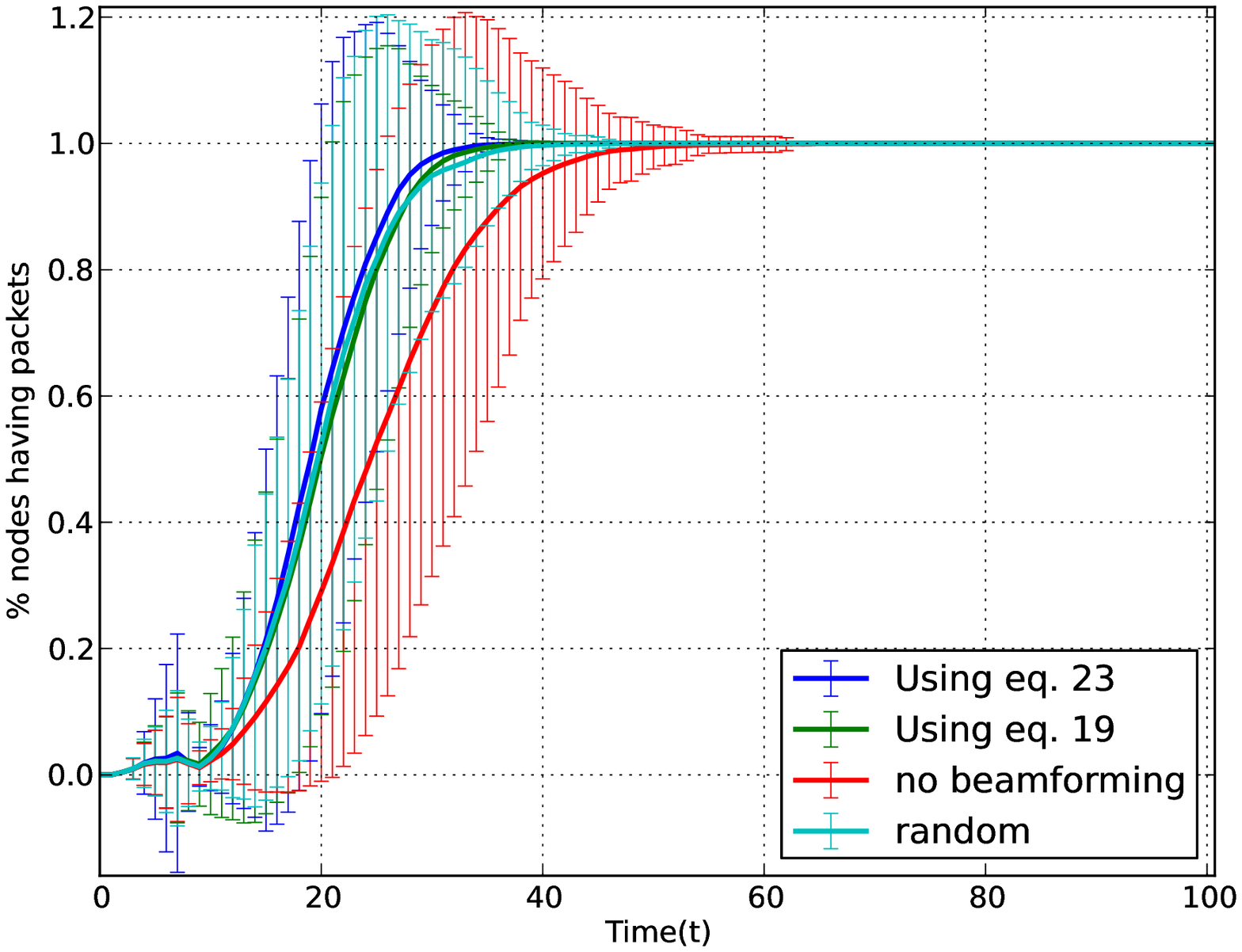}
                \label{subfig:numpacket2}
            }
        }
        \mbox
        {
            \hspace{-0.2cm}
            \subfigure[\% nodes having packets when single node acts as a source and it has multiple packets to transmit.]
            {
                \includegraphics[width = 0.45\textwidth]{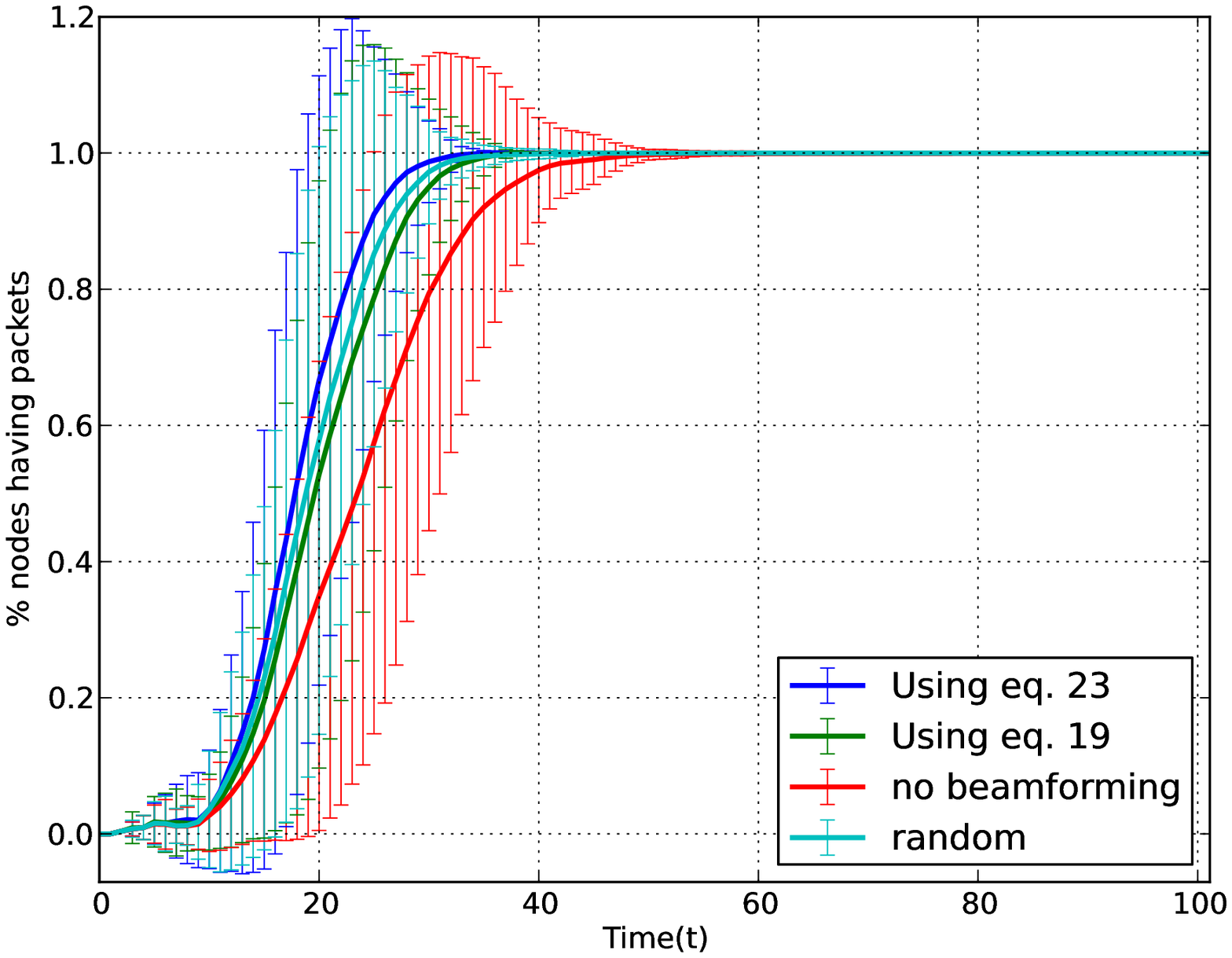}
                \label{subfig:numpacket3}
            }
            \hspace{1cm}
            \subfigure[\% nodes having packets when multiple nodes act as source and each of them has single packet to transmit.]
            {
                \includegraphics[width = 0.45\textwidth]{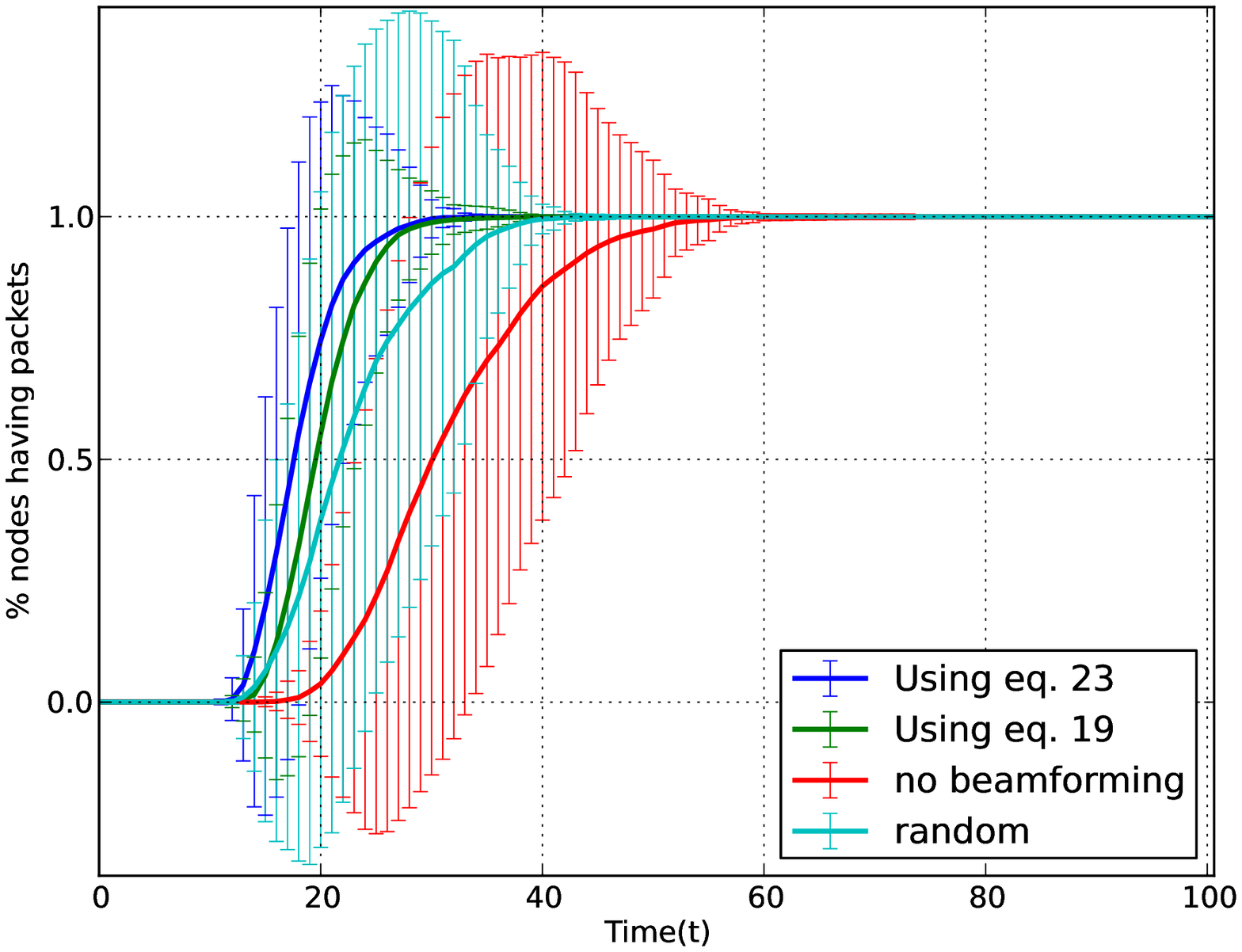}
                \label{subfig:numpacket4}
            }
        }
        \mbox
        {
            \hspace{-0.2cm}
            \subfigure[\% nodes having packets when multiple nodes act as source and each of them have multiple packets to transmit.]
            {
                \includegraphics[width = 0.45\textwidth]{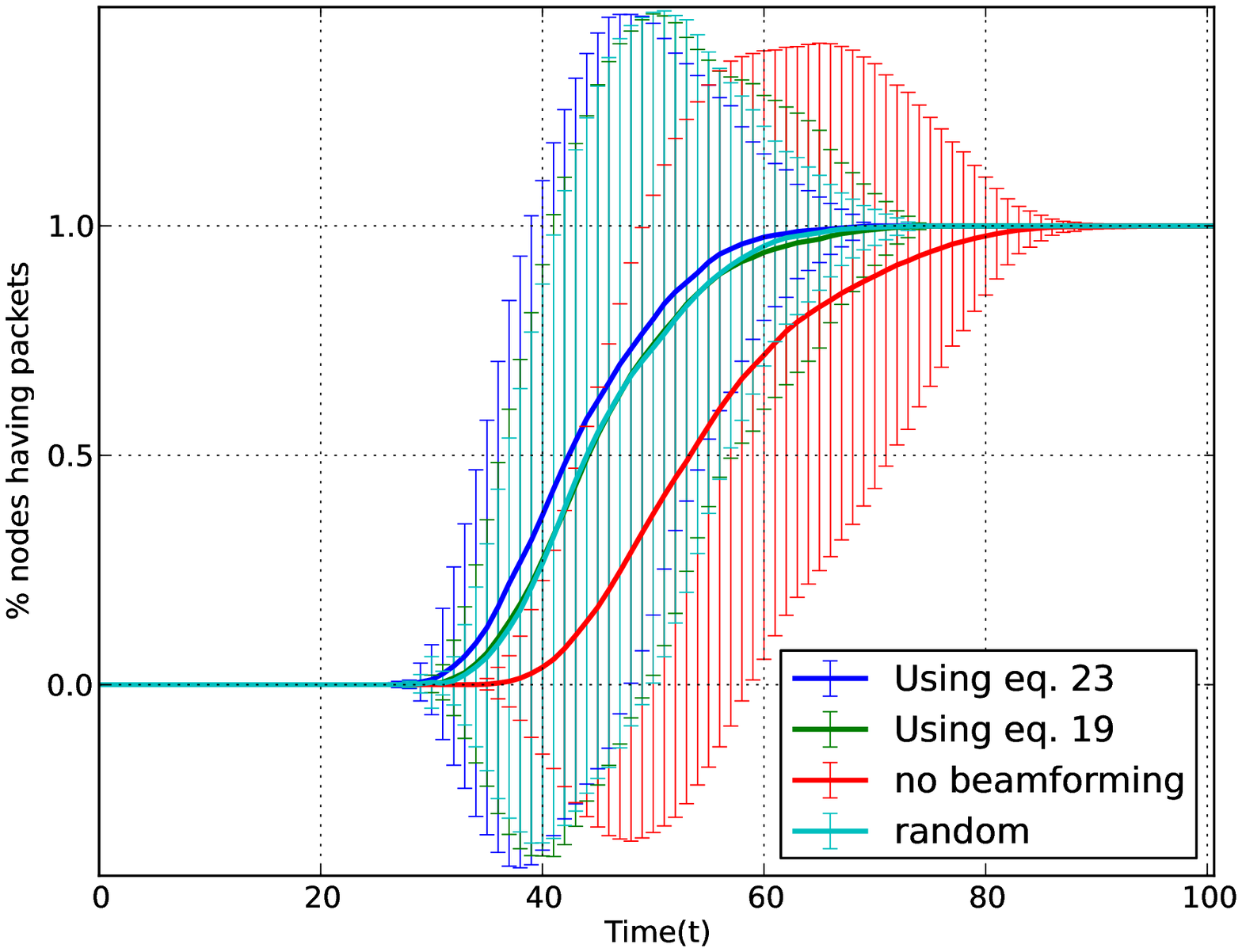}
                \label{subfig:numpacket5}
            }
            \hspace{1cm}
            \subfigure[\% nodes having packets when multiple nodes act as source and each of them have multiple packets to transmit. Sources can join as time passes by.]
            {
                \includegraphics[width = 0.45\textwidth]{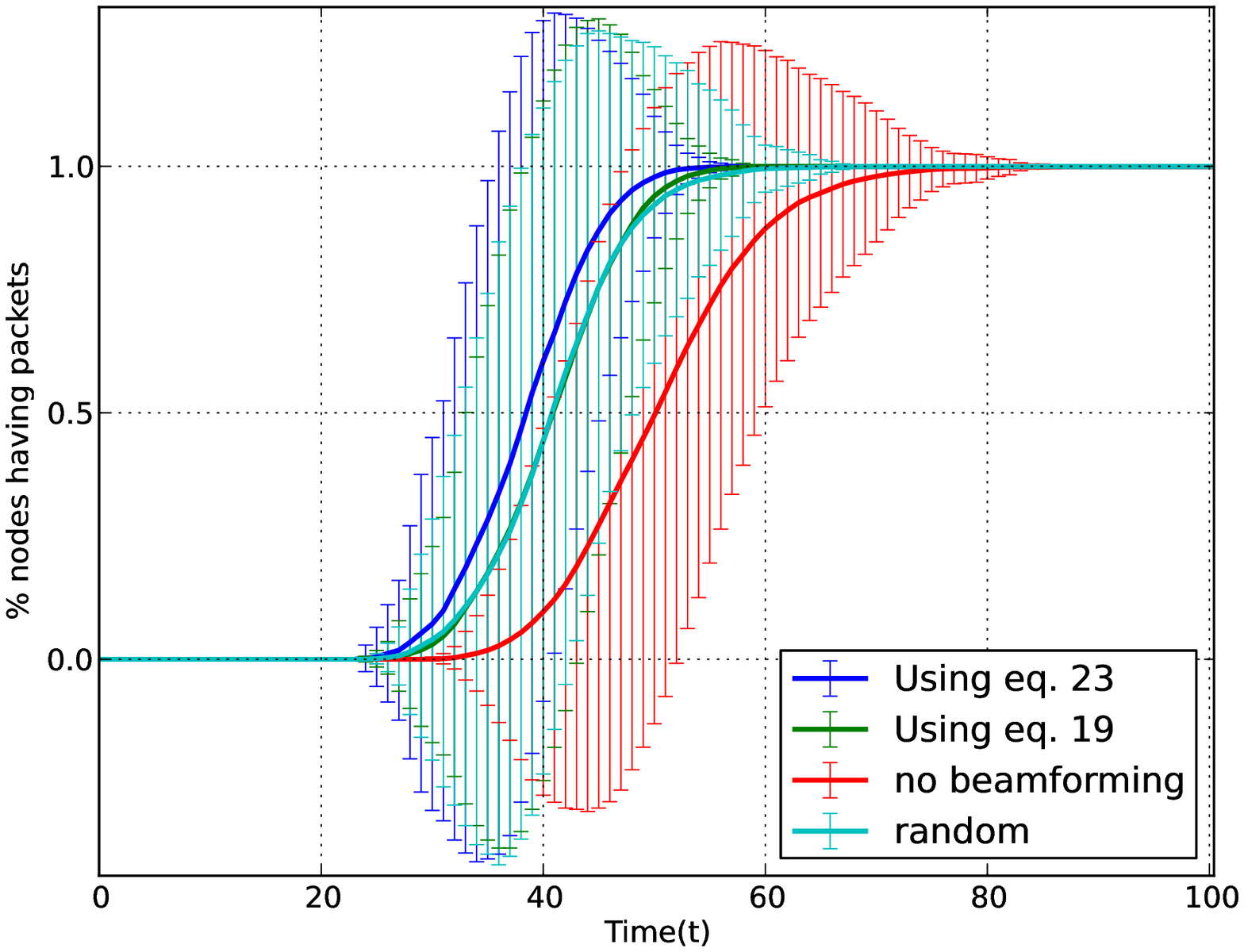}
                \label{subfig:numpacket5_1}
            }
        }
        \caption{Percentage of nodes having packets.}
        \label{fig:results1}
    \end{figure*}

    We first show results obtained for the case when there is only one source having only one packet to transmit. In this scenario, the packet is generated at the start of the simulation in this case. Figure \ref{subfig:numpacket1} shows an improvement achieved when our model is used over the case when there was no beamforming done. The results show that in almost 26 time steps all nodes in the network have the packet when beamforming is done using our model while only $86\%$ nodes have the packet after 26 time steps when no beamforming is done. The results show high variations because the source node is randomly chosen in the network. In each run, due to difference in the $APL$ of the source node to other nodes, each run results into different number of nodes receiving packets. The results also show improvements over the other two validation models ($96\%$ nodes have packets after 26 time steps in both the two other validation models). Further, the results also prove that higher performance is achieved with less energy consumed when compared to model in section \ref{subsubsec:neighborhood based}.

    Figure \ref{subfig:numpacket2} also shows an improvement achieved when there is only single source having updated packet. The initial packet is generated at the start of the simulation in this case while the updated packets are generated until $10$ initial time steps. This is done so that percolation could be observed. Here, one packet is generated in one time step with the packet generated at 10th time step being the most updated packet. Here time take to completely disseminate the updated packets is shifted by $\approx10$ time steps than the previous case because of the updated packets being generated until 10 time steps. All other observations in number of nodes having the packets remain same to that of the previous case. Further, in this case there is a hill between time step 0 and 10 because the updated packets are not generated at all time steps between 0 and 10. When at a time step an updated packet is not generated, the network assumes that there is no updated packet in the network. Under same settings, similar results to that of fig. \ref{subfig:numpacket2}, are obtained for the case when there is only single source which has multiple packets to disseminate in the network (CF. fig. \ref{subfig:numpacket3}).

    Figure \ref{subfig:numpacket4} also shows an improvement achieved when there are multiple sources which have single packet to disseminate in the network. We set number of sources to be 40 for all the scenarios dealing with multiple sources. Similar to the case of single source, the initial packet is generated at the start of the simulation by all the sources. The results show that in $\approx38$ time steps all packets can be disseminated in the network when beamforming is used, however, only $\approx80\%$ nodes have received all the packets when there was no beamforming (CF. fig \ref{subfig:numpacket4}). Further, as there are multiple sources, the number of nodes having all the packets has a shift in time. It takes $\approx11$ time steps for one node to have all the packets from all the sources for the beamforming case while $\approx16$ time steps when not beamforming.

    Figure \ref{subfig:numpacket5} also shows an improvement achieved when there are multiple sources with multiple packets to disseminate in the network. Each source randomly generates packets until $30$ time steps. This makes each source to have different number of packets to be transmitted in the network. This is done so that percolation could be observed. The results show that in $\approx65$ time steps all packets can be disseminated in the network when beamforming is used, however, only $\approx80\%$ nodes have received all the packets when there was no beamforming (CF. fig \ref{subfig:numpacket5}). Further, as there are multiple sources having multiple packets, the number of nodes having all the packets has a shift in time when compared to the previous case. This shift is almost by $\approx16$ time steps before dissemination takes off.

    Figure \ref{subfig:numpacket5_1} shows an improvement achieved when there are multiple sources with multiple packets to disseminate in the network where sources can join as time passes by. Sources can join only until $20$ time steps so that percolation of the number of nodes receiving the packets can be observed. Each source randomly generates packets until $30$ time steps. This makes each source to have different number of packets to be transmitted in the network as discussed above. The results show that in $\approx56$ time steps all packets can be disseminated in the network when beamforming is used, however, only $\approx75\%$ nodes have received all the packets when there was no beamforming (CF. fig \ref{subfig:numpacket5}). Compared to the results obtained for multiple source single packet case, (Cf. fig. \ref{subfig:numpacket4}), there is a shift of almost $\approx13$ time steps before dissemination takes off for the beamforming case.

\begin{figure*}
        \centering
        \mbox
        {
            \subfigure[\% of nodes having packets for variable $\rho$.]
            {
                \includegraphics[width = 0.38\textwidth]{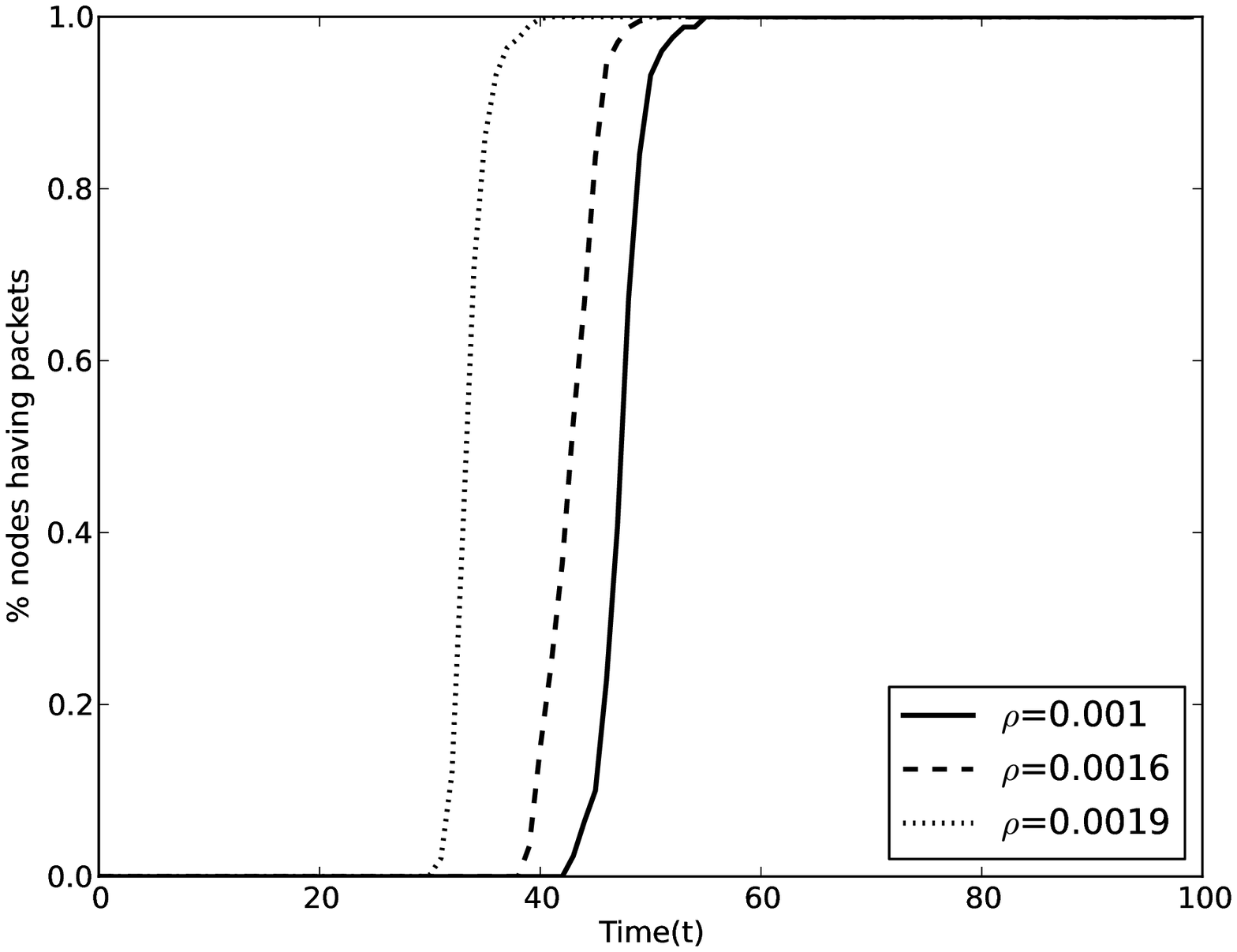}
                \label{subfig:numpacket6}
            }\hspace{1cm}
            \subfigure[\% of nodes having packets for variable $\alpha$.]
            {
                \includegraphics[width = 0.38\textwidth]{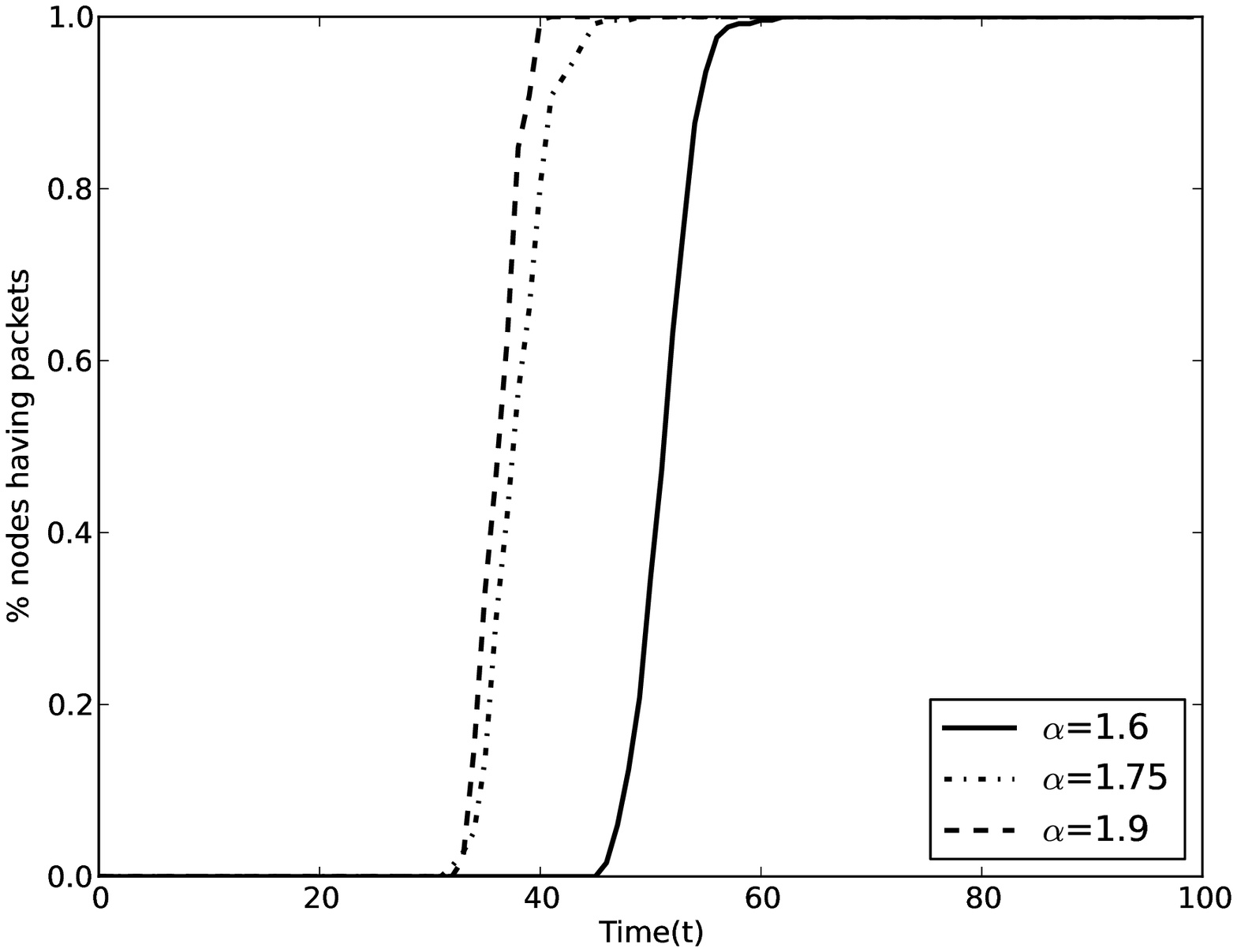}
                \label{subfig:numpacket7}
            }
        }
        \mbox
        {
            \subfigure[\% of nodes having packets for variable $\beta$.]
            {
                \includegraphics[width = 0.38\textwidth]{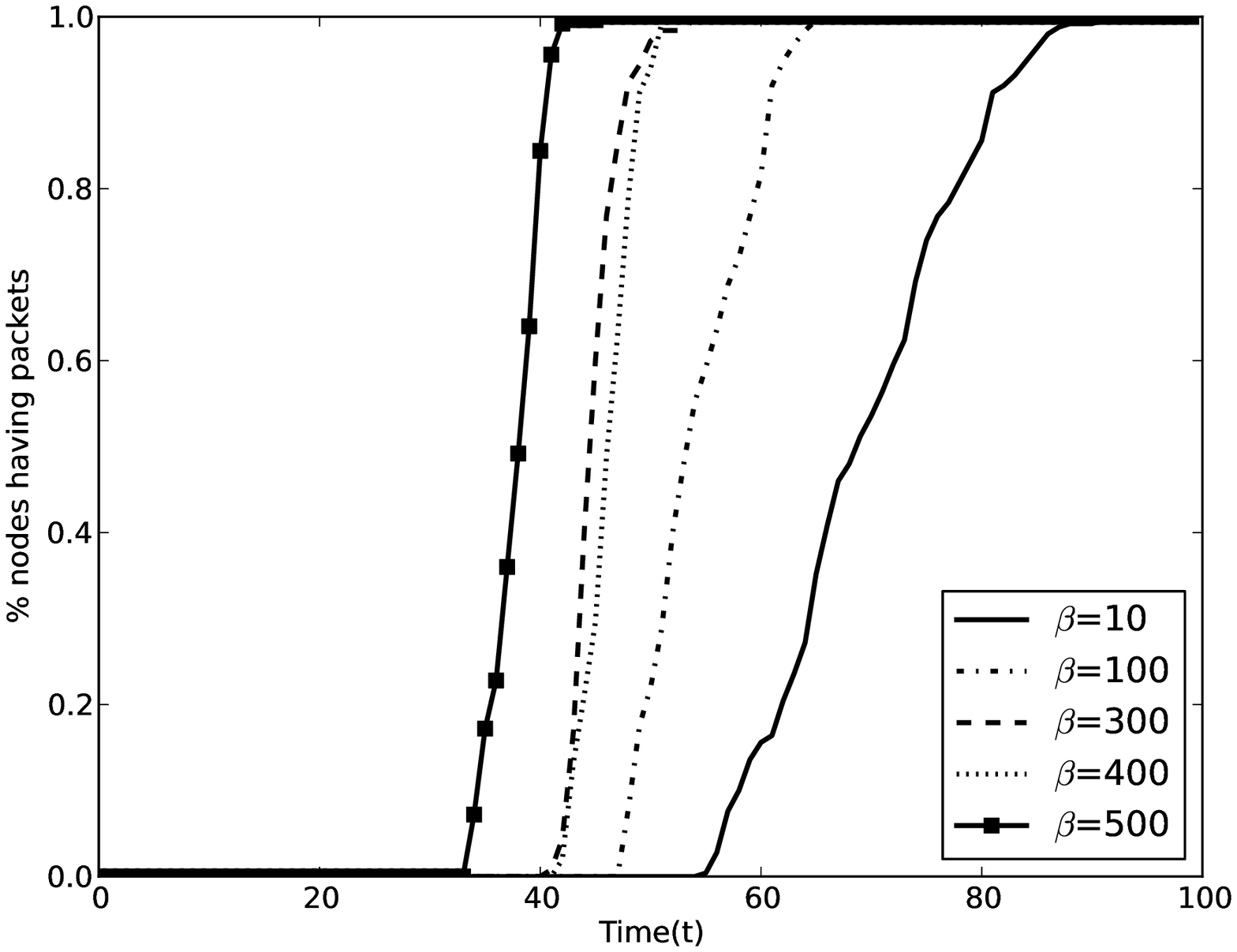}
                \label{subfig:numpacket8}
            }\hspace{1cm}
            \subfigure[\% of nodes having packets for variable $r$.]
            {
                \includegraphics[width = 0.38\textwidth]{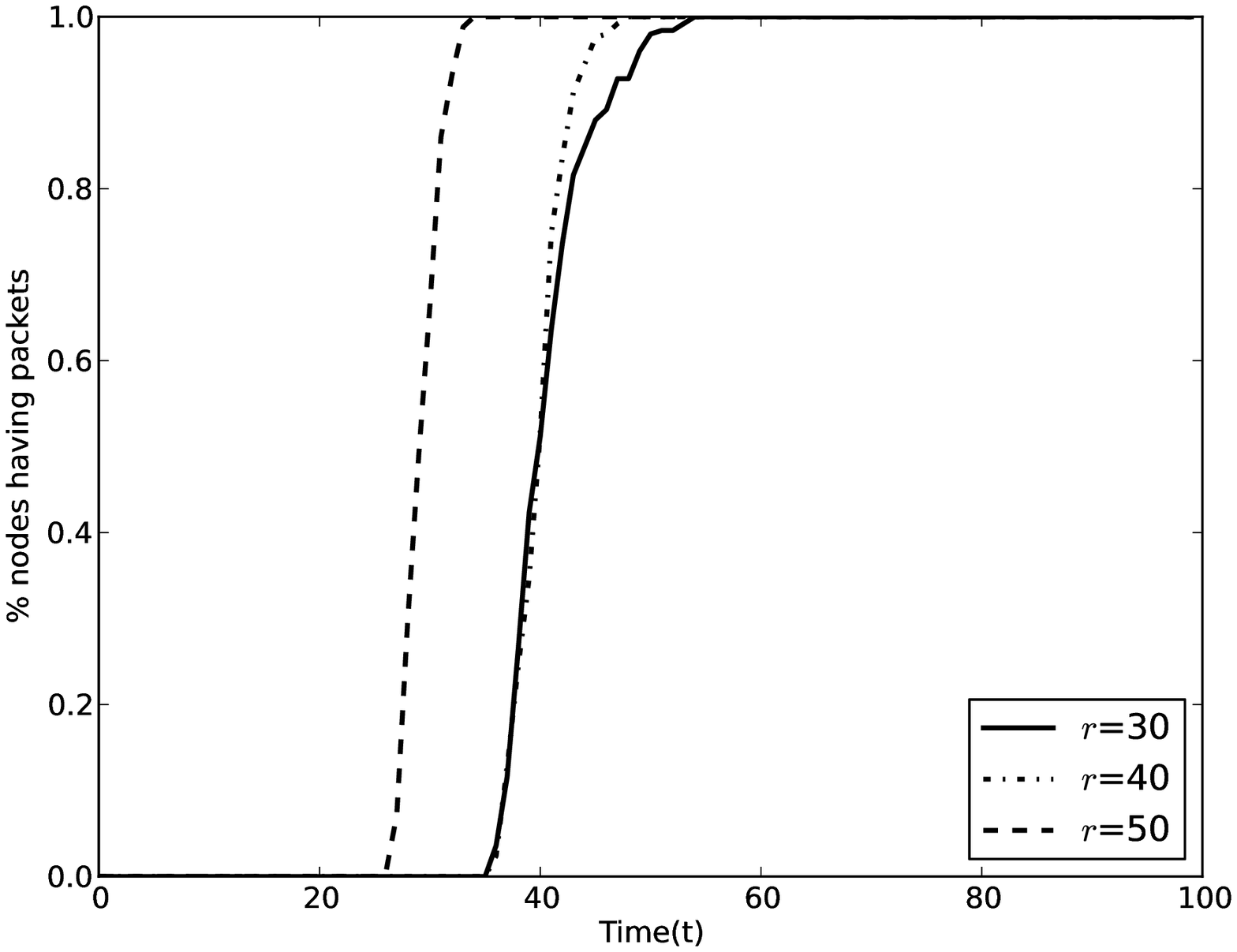}
                \label{subfig:numpacket9}
            }
        }
        \mbox
        {
            \subfigure[\% of nodes having packets for variable $M$.]
            {
                \includegraphics[width = 0.38\textwidth]{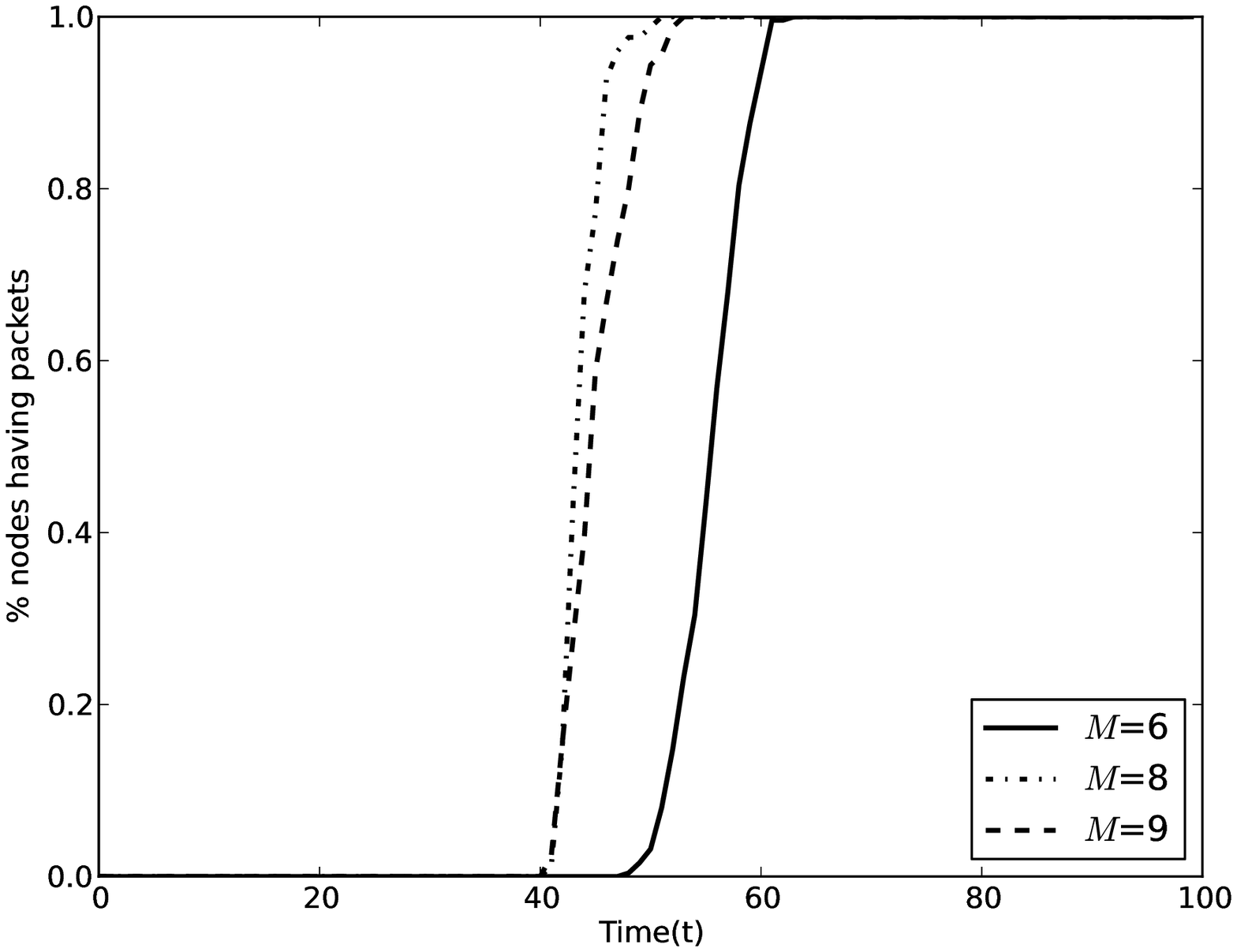}
                \label{subfig:numpacket10}
            }\hspace{1cm}
            \subfigure[\% of nodes having packets for variable $S_{min}$.]
            {
                \includegraphics[width = 0.38\textwidth]{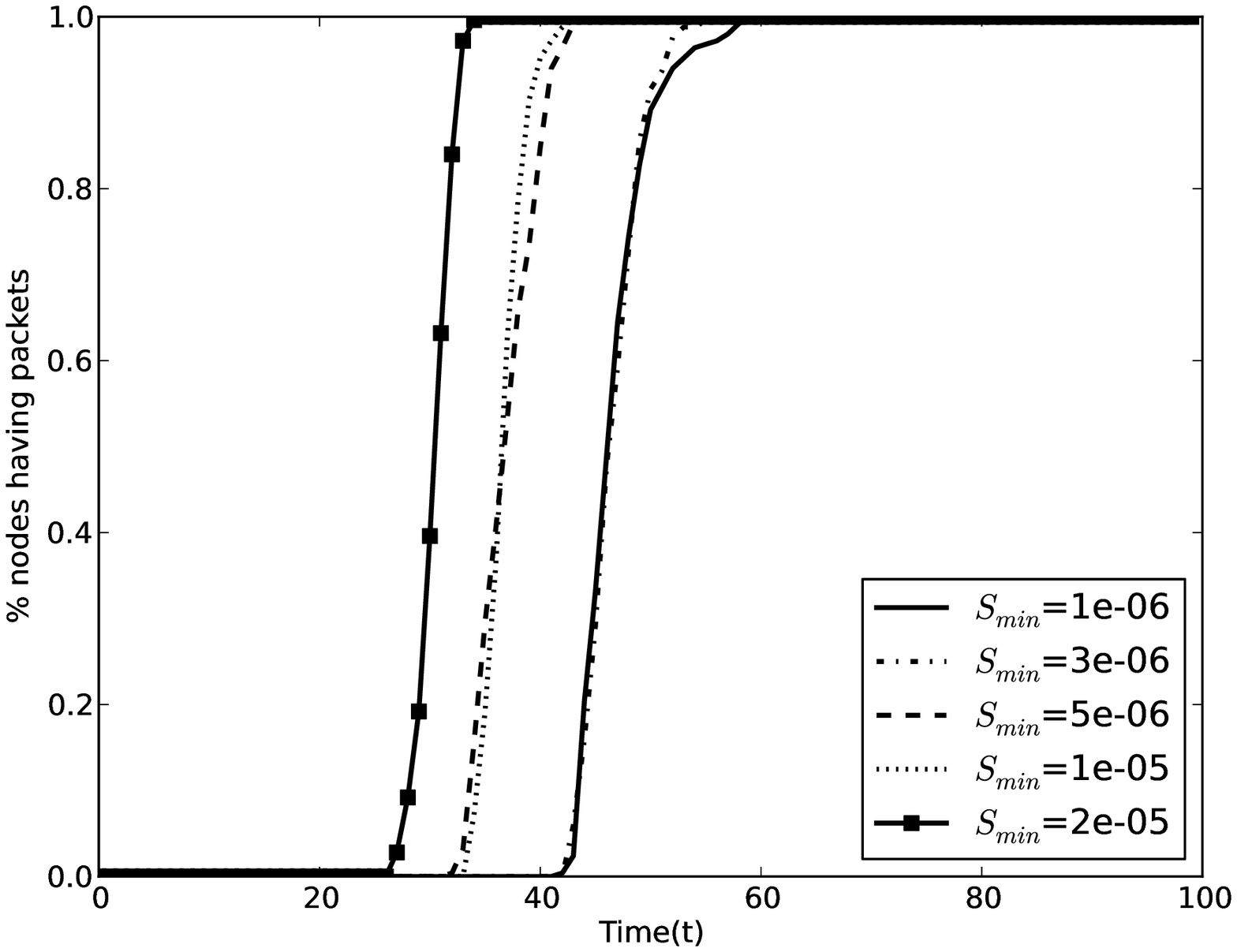}
                \label{subfig:numpacket11}
            }
        }
        \mbox
        {
            \subfigure[\% of nodes having packets for variable $S_{max}$.]
            {
                \includegraphics[width = 0.38\textwidth]{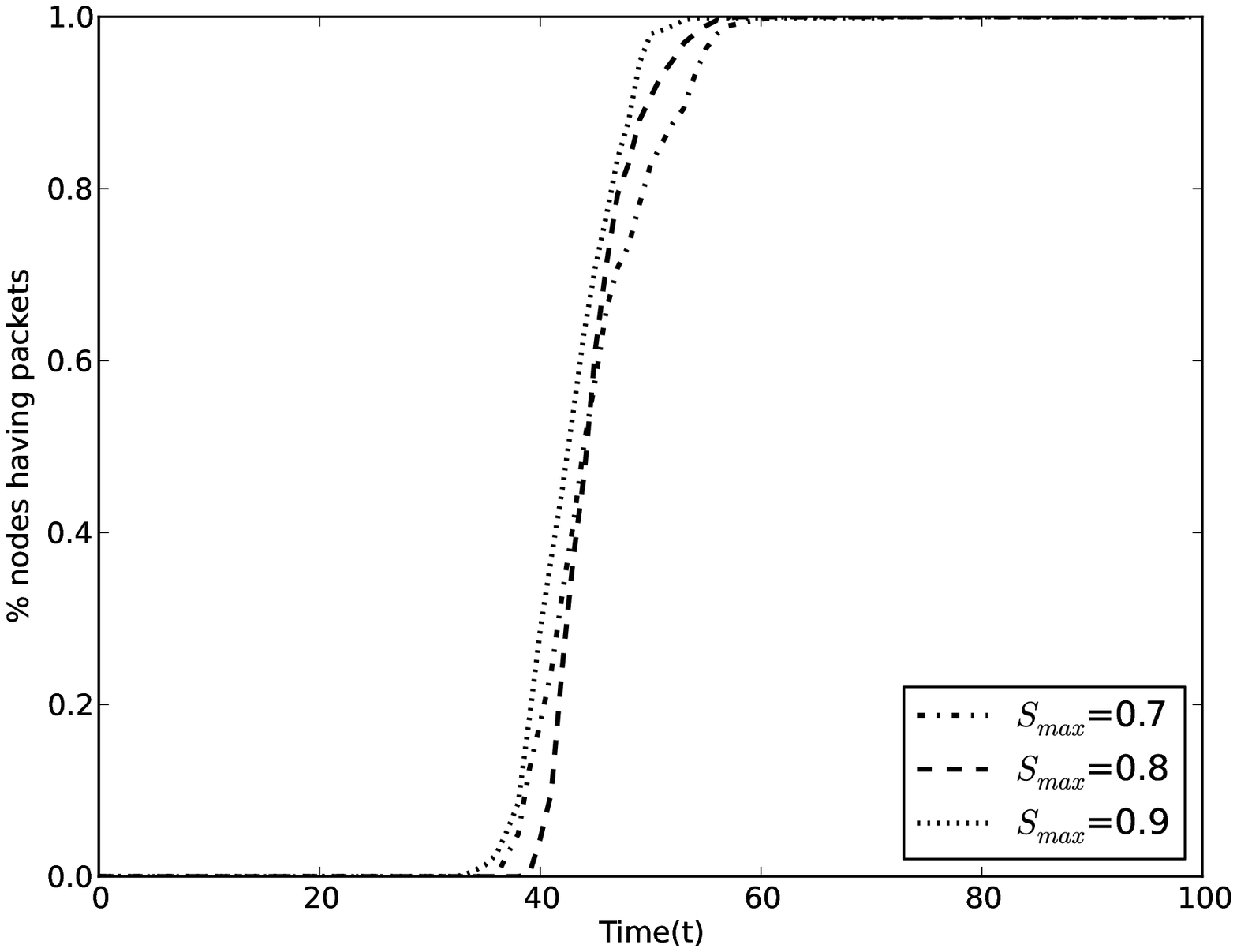}
                \label{subfig:numpacket12}
            }\hspace{1cm}
            \subfigure[\% of nodes having packets for different antennas.]
            {
                \includegraphics[width = 0.38\textwidth]{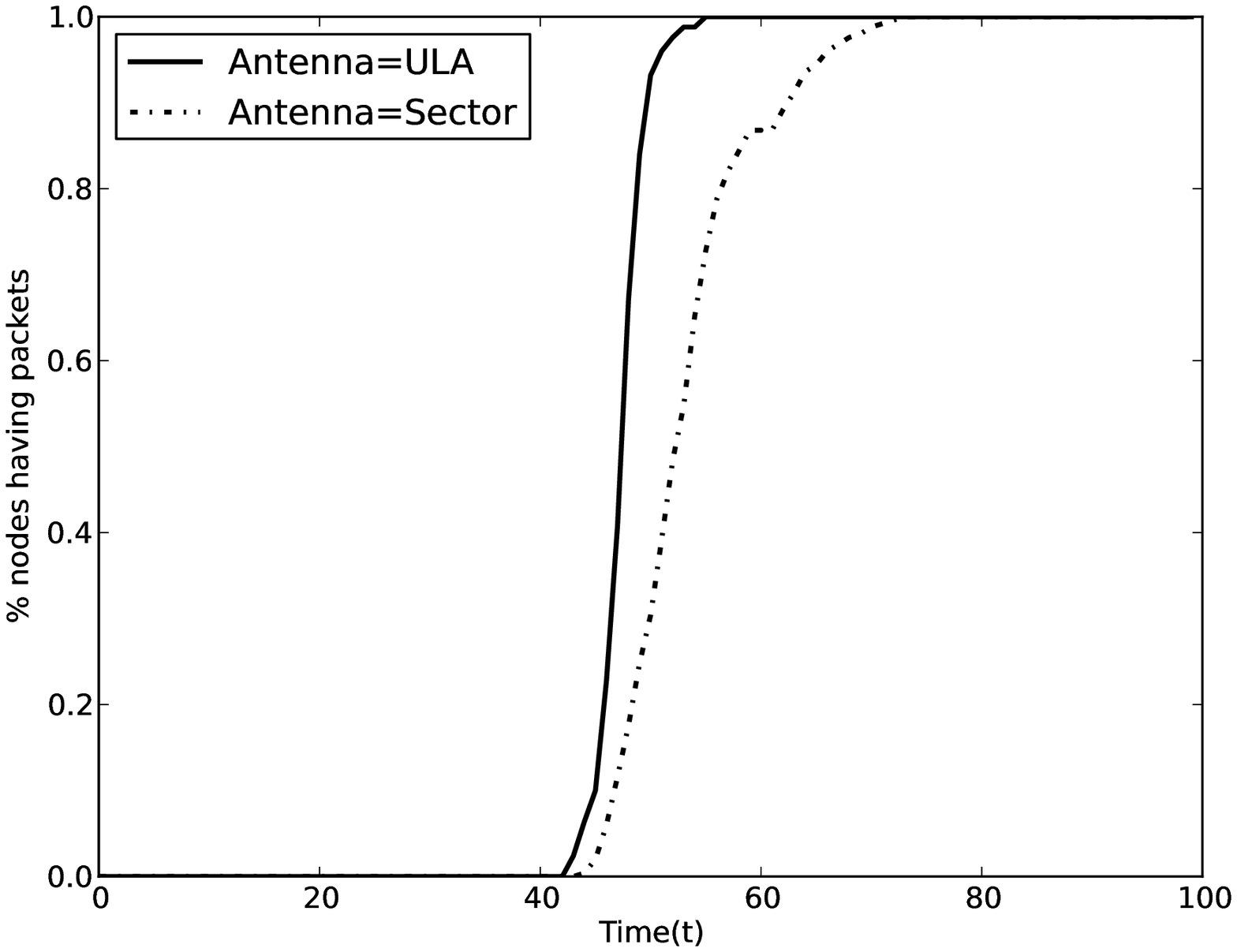}
                \label{subfig:numpacket14}
            }
        }
        \caption{Percentage of nodes having packets when multiple nodes act as source and each of them have multiple packets to transmit.}
        \label{fig:results2}
    \end{figure*}

    We next provide results for varying $\rho$, $\alpha$, $\beta$, $r$, $M$, $S_{min}$, $S_{max}$ and antenna type for the case when there are multiple sources with multiple packets to transmit. In these set of simulations we show the effect of the above mentioned parameters on information dissemination when our model is used. First we show the affect of varying $\rho$ on the model. As $\rho$ increases the number of time steps taken to disseminate the packet reduces. This is because when $\rho$ is high more nodes are in the giant component and the connectivity is high, (Cf. fig \ref{subfig:numpacket6}).

    For varying $\alpha$ it was observed that when $\alpha$ is high the time taken to spread the information in the network decreases, (Cf. fig. \ref{subfig:numpacket7}). For varying $\beta$ variations in time take for all the nodes to receive the information is shown in fig. \ref{subfig:numpacket8}. For less $\beta$, when the jump length cutoff is less, the nodes are not able to move far. This affects the nodes meeting different nodes in the network, thus more time is required to spread the information in the network. While when $\beta$ is more, that is when jump length cutoff is high, nodes are able to move far and meet more nodes and spread the information in the network faster.

    Next we show the results obtained when $r$ and $M$ varies. Results clearly state that when $r$ is high more nodes are connected and the dissemination of packets is faster, (CF. fig. \ref{subfig:numpacket9}). For varying $M$ the results state that when $M$ is high the information dissemination is much faster, (CF. fig. \ref{subfig:numpacket10}). This behavior can be explained as when $M$ is high the beam length is also high and thus more nodes can be connected.

    Variations in $S_{min}$ and $S_{max}$ affects the number of nodes beamforming and to whom they beamform. Low value of $S_{min}$ reduces the number of nodes beamforming while a high value increases that number. This affects the information dissemination in the network, (CF. fig \ref{subfig:numpacket11}). However, low value of $S_{max}$ increases the number of nodes the nodes can beamform to while a high value decreases that number. Moreover, as defined in table \ref{table:beamformer} there can be other nodes to which beams can be directed that do not have high stability. A high value of $S_{max}$ only reduces the weight of a sector defined by eq. \ref{eq:bd} by some fraction.  As one sector that has the best weight has to be chosen, $S_{max}$ should not affect the spreading of the information in the network, (CF. fig \ref{subfig:numpacket12}).

    Next we show results obtained when different types of antennas are used. We used two different types of antennas, namely, $ULA$ and a theoretical antenna model (Sector Model). $ULA$ achieves better results than sector model because in $ULA$ there is a secondary lobe with substantial beam length and width, (CF. fig \ref{subfig:numpacket14}). This increases the connectivity of the node as compared to when sector model is used.

    \begin{table}
        \centering
        \begin{tabular}{|l|l|l|}
            \hline
            \multicolumn{2}{|c|}{\textbf{Parameter}}&\textbf{Time taken}\\
            \hline
            &\textbf{value}&\textbf{value}\\
            \hline
            $\rho$ & $\uparrow$ & $\downarrow$ \\
            $\alpha$ &$\uparrow$& $\downarrow$\\
            $\beta$ & $\uparrow$&$\downarrow$\\
            $r$ & $\uparrow$&$\downarrow$\\
            $m$ & $\uparrow$&$\downarrow$\\
            $S_{min}$ & $\uparrow$ & $\downarrow$\\
            $S_{max}$ & $\uparrow$&$\leftrightarrow$\\
            Number of sources & $\uparrow$ & $\uparrow$\\
            Number of packets & $\uparrow$ & $\uparrow$\\
            \hline
            \textbf{Other Parameters}&&\\
            \hline
            Location of sources &&\\
            Stability measure type &&\\
            Type of antenna array &&\\
            \hline
        \end{tabular}
        \caption{Factors affecting dissemination of packet. The table shows the effect on the time taken to disseminate information in the network completely by the increase in the parameter value.}
        \label{table:parameteraffect}
    \end{table}

    These results led to tabulate table \ref{table:parameteraffect} that summarizes the effect of each parameter, described above, on the information dissemination.

\section{Conclusion}\label{sec:conclusion}
From this paper, we have tried to propose a model that could enhance information dissemination in the mobile scenarios. The contribution of this paper is four fold. First, while doing a literature survey we proposed an enhancement to Zayani et al's. algorithm. Second, we proposed a probabilistic approach to compute node stability in a mobile environment. Third, based on the stability measures we perform beamforming to enhance information dissemination in mobile environment. Fourth, the operations are fully distributed and does not require any global knowledge of the network.

Our results show an improvement in information dissemination when a stability measure is used to beamform over three different cases including the case when no beamforming was done and when randomly nodes were chosen to beamform. Through our results we have also tried to prove that probabilistic calculation of stability can very well enhance the information dissemination in the network and have validated the results. The probabilistic calculation of stability does not require past snapshots but one previous one. Further, our algorithm works in fully decentralized manner. The results show that information dissemination is affected by number of sources, number of packets, mobility parameters, density, radius of communication, number of antennas, antenna type, stability parameters and stability type used.
However, in the current work only fixed number of nodes is considered in the simulation area. In real scenarios, nodes can join as well as can move out of the simulation area. We would like to address these issues and see how information dissemination is affected due to the dynamic population. Further, we would also like to use real mobility traces to validate our results.

\section{Acknowledgement}\label{sec:acknowledgement}
This project was funded by the grant from University of Pierre and Marie Curie, Paris, France.

\bibliography{thebibliography}
\bibliographystyle{elsarticle-num}

\end{document}